\documentclass[conference]{IEEEtran}

\usepackage[pdftex]{graphicx,color}
\usepackage{latexsym,amssymb,ifthen,cite,array,url,afterpage}

\usepackage[cmex10]{amsmath}
\usepackage{array}
\usepackage{fixltx2e}

\newcommand{\ignore}[1]{}

\newcommand{\Fig}[1]{Figure~\ref{#1}}

\newcommand{\eqdef}{\stackrel{\scriptscriptstyle\bigtriangleup}{=} }

\newcommand{\cond}{\hspace{0.02em}|\hspace{0.08em}}

\newcommand{\cent}[1]{\makebox(0,0){#1}}
\newcommand{\pos}[2]{\makebox(0,0)[#1]{#2}}

\newcommand{\T}{\mathsf{T}}

\newcommand{\R}{\mathbb{R}}

\newcommand{\calN}{\mathcal{N}}
\newcommand{\calK}{\mathcal{K}}

\newcommand{\E}{\operatorname{E}}

\newcommand{\argmax}{\operatornamewithlimits{argmax}}

\newcommand{\dW}{\tilde{W}}
\newcommand{\dxi}{\tilde{\xi}}

\newcommand{\dF}{\tilde{F}}

\makeatletter
\DeclareFontFamily{U}{MnSymbolA}{}
\DeclareSymbolFont{MnSyA}{U}{MnSymbolA}{m}{n}
\DeclareFontShape{U}{MnSymbolA}{m}{n}{
<-6> MnSymbolA5
<6-7> MnSymbolA6
<7-8> MnSymbolA7
<8-9> MnSymbolA8
<9-10> MnSymbolA9
<10-12> MnSymbolA10
<12-> MnSymbolA12}{}
\DeclareMathSymbol{\smallrightarrow}{\mathrel}{MnSyA}{0}
\DeclareMathSymbol{\smallleftarrow}{\mathrel}{MnSyA}{2}
\DeclareMathSymbol{\smallleftrightarrow}{\mathrel}{MnSyA}{16}
\newcommand{\smallrightarrowfill@}{\arrowfill@\relbar\relbar\smallrightarrow}
\newcommand{\smallleftarrowfill@}{\arrowfill@\smallleftarrow\relbar\relbar}
\newcommand{\smallleftrightarrowfill@}
{\arrowfill@\smallleftarrow\relbar\smallrightarrow}
\renewcommand{\overrightarrow}{\mathpalette{\overarrow@\smallrightarrowfill@}}
\renewcommand{\overleftarrow}{\mathpalette{\overarrow@\smallleftarrowfill@}}
\renewcommand{\overleftrightarrow}
{\mathpalette{\overarrow@\smallleftrightarrowfill@}}
\makeatother

\providecommand{\msgf}[2]{\protect\overrightarrow{#1}_{\mspace{-3mu}#2}} 
\providecommand{\msgb}[2]{\protect\overleftarrow{#1}_{\mspace{-3mu}#2}} 

\definecolor{gray}{rgb}{0.5, 0.5, 0.5}
\newcommand{\gray}{\color{gray}}

\newenvironment{proofof}[1]{\begin{trivlist}\item[]{\bfseries Proof\ifthenelse{\equal{#1}{}}{}{ #1}:}
 }{\hfill$\Box$\end{trivlist}}

\newcommand{\eproofnegspace}{\\[-1.5\baselineskip]\rule{0em}{0ex}}

\newcounter{examplecntr}
\renewcommand{\theexamplecntr}{\arabic{examplecntr}}
{\begin{trivlist}\small\item[]\refstepcounter{examplecntr}%
 {\bfseries Example~\theexamplecntr%
  \ifthenelse{\equal{#1}{}}{ }{ (#1) } 
}}%
{\hfill$\Box$\end{trivlist}}

\newcounter{theoremcntr}
{\begin{trivlist}\item[]\refstepcounter{theoremcntr}%
{\bfseries Theorem~\thetheoremcntr%
  \ifthenelse{\equal{#1}{}}{}{ (#1)}.
}}%
{\hfill$\Box$\end{trivlist}}

\newcounter{saveequationcntr}

\newcommand{\Beginlocaleqncntr}{%
\setcounter{saveequationcntr}{\value{equation}}%
\setcounter{equation}{0}%
\renewcommand{\theequation}{\Roman{table}.\arabic{equation}}%
}
\newcommand{\Endlocaleqncntr}{\setcounter{equation}{\value{saveequationcntr}}}

\newcommand{\knownBox}{\cent{\rule{1.75\unitlength}{1.75\unitlength}}}

\begin{document}

\title{On Sparsity by NUV-EM, Gaussian Message Passing, and Kalman Smoothing}

\author{\IEEEauthorblockN{Hans-Andrea Loeliger, Lukas Bruderer, Hampus Malmberg, Federico Wadehn, and Nour Zalmai}
\IEEEauthorblockA{ETH Zurich, 
Dept.\ of Information Technology \& Electrical Engineering
}
}

\maketitle

\begin{abstract}
Normal priors with unknown variance (NUV) have long been known
to promote sparsity and to blend well with parameter learning by 
expectation maximization (EM).
In this paper, we advocate this approach for linear state space models
for applications such as the estimation of impulsive signals,
the detection of localized events, 
smoothing with occasional jumps in the state space, 
and the detection and removal of outliers.

The actual computations boil down to multivariate-Gaussian
message passing algorithms that are closely related to Kalman smoothing. 
We give improved tables of Gaussian-message computations
from which such algorithms are easily synthesized,
and we point out two preferred such algorithms.
\end{abstract}
\vspace{1ex}

\section{Introduction}

This paper is about two topics:
\begin{enumerate}
\item
A particular approach to modeling and estimating sparse parameters
based on zero-mean normal priors with unknown variance (NUV).
\item
Multivariate-Gaussian message passing ($\approx$ variations of Kalman smoothing) in such models.
\end{enumerate}
The main point of the paper is that these two things go 
very well together and combine to a versatile toolbox.
This is not entirely new, of course, and the body of related literature 
is large. 
Nonetheless, the specific perspective of this paper has not, 
as far as known to these authors, been advocated before.

Concerning the second topic, linear state space models 
continue to be an essential tool 
for a broad variety of applications, cf.\ \cite{KSH:LinEs2000b,RoGha:unifLGM1999,DuKa:tsassm,Bi:PRML}.
The primary algorithms for such models are variations and generalizations
of Kalman filtering and smoothing, or, equivalently,
multivariate-Gaussian message passing in the corresponding factor graph 
\cite{Lg:ifg2004,LDHKLK:fgsp2007}
(or similar graphical model \cite{RoGha:unifLGM1999}). 
A variety of such algorithms can easily be synthesized 
from tables of message computations as in \cite{LDHKLK:fgsp2007}. 
In this paper, we give a new version of these tables 
with many improvements over those in \cite{LDHKLK:fgsp2007},
and we point out two preferred such algorithms.

Concerning the first topic, 
NUV priors (zero-mean normal priors with unknown variance) 
originated in Bayesian inference \cite{McKay:Bi1992,Gull:bii1988,Neal:blnn1996b}. 
The sparsity-promoting nature of such priors 
is the basis of automatic relevance determination (ARD) 
and sparse Bayesian learning 
developed by Neal \cite{Neal:blnn1996b}, Tipping \cite{Tip:RVM2001,Tipp:fmlm2003},
Wipf et al.\ \cite{WiNag:nvARD2007,WiRao:splbs2004},
and others.

The basic properties of NUV priors are illustrated by the following simple example.
Let $U$ be a variable or parameter of interest, which 
we model as a zero-mean real scalar Gaussian random variable
with unknown variance $s^2$. 
Assume that we observe \mbox{$Y=U+Z$}, where the noise $Z$
is zero-mean Gaussian with (known) variance $\sigma^2$ and independent of $U$.
The maximum likelihood (ML) estimate of $s^2$ 
from a single sample $Y=\mu\in \R$
is easily determined:
\begin{IEEEeqnarray}{rCl}
\hat s^2 
  & \eqdef & \argmax_{s^2} 
     \frac{1}{\sqrt{2\pi (s^2 + \sigma^2)}} e^{- \mu^2/2(s^2+\sigma^2)} \label{eqn:BasicScalarExArgmax}\\
  & = & \max\{ 0,\, \mu^2 - \sigma^2 \}.  \label{eqn:BasicScalarSigmaML}
\end{IEEEeqnarray}
In a second step, 
for $s^2$ fixed to $\hat s^2$ as in (\ref{eqn:BasicScalarSigmaML}),
the MAP/MMSE/LMMSE estimate of $U$ is
\begin{IEEEeqnarray}{rCl}
\hat{u} & = & \mu \cdot \frac{\hat s^2}{\hat s^2 + \sigma^2} \\
    & = & \left\{ \begin{array}{ll}
            \mu \cdot \frac{\mu^2-\sigma^2}{\mu^2} & \text{if $\mu^2>\sigma^2$}\\
            0, & \text{otherwise.}
          \end{array}\right.
       \IEEEeqnarraynumspace\label{eqn:BasicScalarUMAP}
\end{IEEEeqnarray}

Equations (\ref{eqn:BasicScalarExArgmax})--(\ref{eqn:BasicScalarUMAP}) 
continue to hold if the scalar observation $Y$ is generalized 
to an observation \mbox{$Y\in \R^N$} such that,
for fixed \mbox{$Y=y$}, the likelihood function $p(y \cond u)$
is Gaussian (up to a scale factor) with mean $\mu$ and variance $\sigma^2$.
In fact, this is all we need to know in this paper about NUV priors per se.

The estimate (\ref{eqn:BasicScalarUMAP}) 
has some pleasing properties: 
first, it promotes sparsity and can thus be used to select features or relevant parameters;
second, it has no \emph{a priori} preference as to the scale of $U$, 
and large values of $U$ are not scaled down.
Note that the latter property is lost if ML estimation of $s^2$ 
is replaced by MAP estimation based on a proper prior on $s^2$.

In this paper, we will stick to basic NUV regularization 
as above,
with no prior on the unknown variances: 
variables or parameters of interest are modeled as 
independent Gaussian random variables,
each with its own unknown variance that is estimated (exactly or approximately) 
by maximum likelihood. 
We will advocate the use of NUV regularization 
in linear state space models, for applications 
such as the estimation of impulsive signals,
the detection of localized events, 
smoothing with occasional jumps in the state space,
and the detection and removal of outliers. 

Concerning the actual computations, estimating the unknown variances
is not substantially different from learning other parameters 
of state space models and can be carried out by 
expectation maximization (EM) \cite{DLR:EM1977,StSe:cmem2004,DEKL:em2013arXiv,GhaHi:pelds1996}
and other methods
in such a way that the actual computations 
essentially amount to Gaussian message passing.

The paper is structured as follows. 
In Section~\ref{sec:LeastSquares}, we begin with a quick look at NUV regularization 
in a standard linear model. 
Estimation of the unknown variances is addressed in Section~\ref{sec:VarEst}.
Factor graphs and state space models are reviewed 
in Sections \ref{sec:FG} and \ref{sec:StateSpaceModels}, respectively,
and NUV regularization in such models is addressed in Section~\ref{sec:StateSpaceNUV}.
The new tables of Gaussian-message computations are given in 
Appendix~\ref{appsec:GMPTables}.

\begin{figure*}
\centering
\begin{picture}(115,55)(0,-2)

\put(12,42){\framebox(6,6){$\calN$}}
 \put(15,42){\vector(0,-1){7}}
\put(12.5,30){\framebox(5,5){$\times$}}
 \put(2.5,32.5){\vector(1,0){10}}   \put(6,33.5){\pos{bc}{$\sigma_1$}}
\put(15,30){\vector(0,-1){9.5}}    \put(14.25,25.5){\pos{cr}{$U_1$}}
\put(12,14.5){\framebox(6,6){$b_1$}}
\put(15,14.5){\vector(0,-1){9.5}}  \put(14.25,10){\pos{cr}{$\tilde U_1$}}

\put(37,42){\framebox(6,6){$\calN$}}
 \put(40,42){\vector(0,-1){7}}
\put(37.5,30){\framebox(5,5){$\times$}}
 \put(27.5,32.5){\vector(1,0){10}}  \put(31,33.5){\pos{bc}{$\sigma_2$}}
\put(40,30){\vector(0,-1){9.5}}     \put(39.25,25.5){\pos{cr}{$U_2$}}
\put(37,14.5){\framebox(6,6){$b_2$}}
\put(40,14.5){\vector(0,-1){9.5}}   \put(39.25,10){\pos{cr}{$\tilde U_2$}}

\put(57.5,17.5){\cent{\ldots}}

\put(72,42){\framebox(6,6){$\calN$}}
 \put(75,42){\vector(0,-1){7}}
\put(72.5,30){\framebox(5,5){$\times$}}
 \put(62.5,32.5){\vector(1,0){10}}  \put(66,33.5){\pos{bc}{$\sigma_K$}}
\put(75,30){\vector(0,-1){9.5}}     \put(74.25,25.5){\pos{cr}{$U_K$}}
\put(72,14.5){\framebox(6,6){$b_K$}}
\put(75,14.5){\vector(0,-1){9.5}}   \put(74.25,10){\pos{cr}{$\tilde U_K$}}

\put(97,14.5){\framebox(6,6){}}     \put(100,21.5){\pos{cb}{$\calN(0,\sigma^2 I)$}}
\put(100,14.5){\vector(0,-1){9.5}}  \put(99,10){\pos{cr}{$Z$}}

\put(0,2.5){\knownBox}              
                                    \put(0,0.5){\pos{ct}{$X_0=0$}}
\put(0,2.5){\vector(1,0){12.5}}     
\put(12.5,0){\framebox(5,5){$+$}}
\put(17.5,2.5){\vector(1,0){20}}    \put(27.5,1){\pos{ct}{$X_1$}}
\put(37.5,0){\framebox(5,5){$+$}}
\put(42.5,2.5){\line(1,0){7.5}}
\put(57.5,2.5){\cent{\ldots}}
\put(62.5,2.5){\vector(1,0){10}}
\put(72.5,0){\framebox(5,5){$+$}}
\put(77.5,2.5){\vector(1,0){20}}    \put(87.5,1){\pos{ct}{$X_K$}}
\put(97.5,0){\framebox(5,5){$+$}}
\put(102.5,2.5){\line(1,0){12.5}}
 \put(102.5,2.5){\vector(1,0){11.625}}
\put(115,2.5){\knownBox}            \put(113,0.5){\pos{ct}{$Y=y$}}

\end{picture}
\vspace{2mm}
\caption{\label{fig:FGCS}%
Cycle-free factor graph of (\ref{eqn:sumOfNUV}) with NUV regularization.}
\end{figure*}
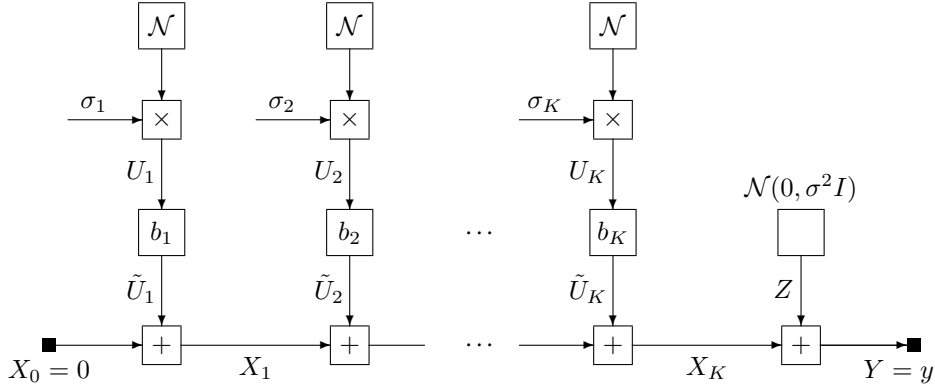

\section{Sum of Gaussians and Least Squares\\ with NUV Regularization}
\label{sec:LeastSquares}

We begin with an elementary linear model 
(a special case of a relevance vector machine \cite{Tip:RVM2001})
as follows.  
For $b_1,\ldots,b_K \in \R^n \setminus \{ 0 \}$, let 
\begin{equation} \label{eqn:sumOfNUV}
Y = \sum_{k=1}^K b_k U_k + Z
\end{equation}
where $U_1,\ldots,U_K$ are independent zero-mean real scalar Gaussian random variables 
with unknown variances $\sigma^2_1,\ldots,\sigma^2_K$,
and where the ``noise'' $Z$ is $\R^n$-valued zero-mean Gaussian 
with covariance matrix $\sigma^2 I$
and independent of $U_1,\ldots,U_K$.
For a given observation $Y=y \in\R^n$, 
we wish to estimate, first, $\sigma^2_1,\ldots,\sigma^2_K$
by maximum likelihood, and second, 
$U_1,\ldots,U_K$ (with $\sigma^2_1,\ldots,\sigma^2_K$ fixed).

In the first step, we achieve sparsity: 
if the ML estimate of $\sigma_k^2$ is zero, 
then $U_k=0$ is fixed in the second step.

The second step\,---\,the estimation of $U_1,\ldots,U_K$ for fixed $\sigma^2_1,\ldots,\sigma^2_K$\,---\,%
is a standard Gaussian estimation problem where MAP estimation, MMSE estimation, 
and LMMSE estimation coincide
and amount to minimizing
\begin{equation} \label{eqn:LeastSquaresStatement}
\frac{1}{\sigma^2} \Big\| y - \sum_{k\in \calK^+} b_k u_k \Big\|^2
 + \sum_{k\in\calK^+}  \frac{1}{\sigma_k^2} \| u_k \|^2,
\end{equation}
where $\calK^+$ denotes the set of those indices $k\in \{1,\ldots,K\}$
for which $\sigma^2_k > 0$. 
A closed-form solution of this minimization is
\begin{equation} \label{eqn:MAPEstimate}
\hat{u}_k = \sigma_k^2 b_k^\T \tilde W y
\end{equation}
with
\begin{equation} \label{eqn:LeastSquarestildeW}
\dW \eqdef \left( \sum_{k=1}^K \sigma_k^2 b_k b_k^\T + \sigma^2 I  \right)^{\!-1} \!\!\!,
\end{equation}
as may be obtained from standard least-squares equations 
(see also \cite{Tipp:fmlm2003}).
An alternative proof will be given in Appendix~\ref{appsec:LeastSquaresMessagePassing},
where we also point out how $\dW$ can be computed without a matrix inversion.

In (\ref{eqn:BasicScalarSigmaML}) and (\ref{eqn:BasicScalarUMAP}), 
the estimate is zero if and only if $y^2 \leq \sigma^2$. 
Two different generalizations of this condition to the setting of this section
are given in the following theorem.
Let $p(y,\ldots)$ denote the probability density of $Y$ and any other variables 
according to (\ref{eqn:sumOfNUV}).
\begin{trivlist}\item{\bf Theorem.}
Let $\sigma_1,\ldots,\sigma_K$ be fixed at a local maximum or at a saddle point 
of $p(y \cond \sigma_1^2,\ldots,\sigma_K^2)$. Then
$\sigma_k^2 = 0$ if and only if
\begin{equation} \label{eqn:LeastSquaresTheoremMsgbUZero}
\big( b_k^\T W_k\, y \big)^2 \leq b_k^\T W_k\, b_k
\end{equation}
with
\begin{equation} \label{eqn:DefWk}
W_k \eqdef \left( \sum_{\ell=1}^K \sigma_\ell^2 b_\ell b_\ell^\T + \sigma^2 I - \sigma_k^2 b_k b_k^\T  \right)^{\!-1} \!\!\!.
\end{equation}
Moreover, with $\dW$ as in (\ref{eqn:LeastSquarestildeW}), 
we have
\begin{equation} \label{eqn:LeastSquaresTheoremdWIneq}
\big( b_k^\T \dW y \big)^2 \leq b_k^\T \dW b_k^\T,
\end{equation}
with equality if $\sigma_k^2 >0$.
\hfill$\Box$%
\end{trivlist}
%
(The proof will be given in Appendix~\ref{appsec:LeastSquaresMessagePassing}.)
The matrices $W_k$ 
and $\dW$ 
are both positive definite. The former depends on $k$, but not on $\sigma^2_k$;
the latter depends also on $\sigma_k^2$, but not on $k$.

\section{Variance Estimation}
\label{sec:VarEst}

Following a standard approach,
the unknown variances $\sigma_1^2,\ldots,\sigma_K^2$
in Section~\ref{sec:LeastSquares}
(and analogous quantities in later sections)
can be estimated by an EM algorithm as follows.
\begin{enumerate}
\item
Begin with an initial guess of $\sigma_1^2,\ldots,\sigma_K^2$.
\item
Compute the mean $m_{U_k}$ and the variance $\sigma_{U_k}^2$ 
of the (Gaussian) posterior distribution $p(u_k \cond y, \sigma_1^2,\ldots,\sigma_K^2)$
with $\sigma_1^2,\ldots,\sigma_K^2$ fixed.
\item
Update $\sigma_1^2,\ldots,\sigma_K^2$ according to (\ref{eqn:NewVarBasicEM}) below.
\item
Repeat steps 2 and~3 until convergence, or until some pragmatic stopping criterion is met.
\item
Optionally update $\sigma_1^2,\ldots,\sigma_K^2$ 
according to (\ref{eqn:NewVarMargML}) below.
\end{enumerate}

The standard EM update for the variances is
\begin{IEEEeqnarray}{rCl}
\sigma_k^2 & \leftarrow & \E\!\big[ U_k^2 \cond \sigma_1^2,\ldots,\sigma_K^2 \big] \\
    & = & m_{U_k}^2 + \sigma_{U_k}^2.  \label{eqn:NewVarBasicEM}
        \IEEEeqnarraynumspace
\end{IEEEeqnarray}
The required quantities $m_{U_k}^2$ and $\sigma_{U_k}^2$ 
are given by (\ref{eqn:LeastSquaresPosteriorMean}) 
and (\ref{eqn:LeastSquaresPosteriorVar}), respectively.
With this update, basic EM theory guarantees 
that the likelihood $p(y \cond \sigma_1^2,\ldots,\sigma_K^2)$
cannot decrease (and will normally increase) in step~3 of the algorithm.

The stated EM algorithm is safe, but the convergence can be slow. 
The following alternative update rule, due to MacKay \cite{Tip:RVM2001},
often converges much faster:
\begin{equation}
\sigma_k^2 \leftarrow \frac{m_{U_k}^2}{1 - \sigma_{U_k}^2 / \sigma_k^2}
\end{equation}
However, this alterative update rule comes without 
guarantees; sometimes, it is too agressive 
and the algorithm fails completely.

An individual variance $\sigma_k^2$ can also be estimated by 
a maximum-likelihood step as in (\ref{eqn:BasicScalarSigmaML}):
\begin{IEEEeqnarray}{rCl}
\sigma_k^2 & \leftarrow & \argmax_{\sigma_k^2} p(y \cond \sigma_1^2,\ldots,\sigma_K^2) \\
   & = & \max \big\{ 0, (\msgb{m}{U_k})^2 - \msgb{\sigma}{U_k}^2 \big\},
   \IEEEeqnarraynumspace\label{eqn:NewVarMargML}
\end{IEEEeqnarray}
The mean $\msgb{m}{U_k}$ is given by (\ref{eqn:LeastSquaresMsgbmUdW})
and the variance $\msgb{\sigma}{U_k}^2$ is given by (\ref{eqn:LeastSquaresMsgbSigmadW}).
However, for parallel updates (simultaneously for all $k\in \{ 1,\ldots, K\}$, 
as in step~3 of the algorithm above), 
the rule (\ref{eqn:NewVarMargML}) is normally too agressive and fails.

Later on, the same algorithm will be used 
for estimating parameters or variables in linear state space models.
In this case, we have no useful analytical expressions 
for (the analogs of) $m_{U_k}$ and $\sigma_{U_k}^2$,
but these quantities are easily computed by Gaussian message passing.

\section{On Factor Graphs and\\ Gaussian Message Passing}
\label{sec:FG}

From now on, we will heavily use factor graphs, 
both for reasoning and for describing algorithms. 
We will use factor graphs as in \cite{LDHKLK:fgsp2007,Lg:ifg2004},
where nodes/boxes represent factors and edges represent variables.
(By contrast, factor graphs as in \cite{KFL:fg2001}
have both variable nodes and factor nodes.)

\Fig{fig:FGCS}, for example, 
represents the probability density $p(y,z,u_1,\ldots,u_K \cond \sigma_1,\ldots,\sigma_K)$
of the model (\ref{eqn:sumOfNUV})
with auxiliary variables $\tilde U_k \eqdef b_k U_k$
and $X_k \eqdef X_{k-1} + \tilde U_k$ with $X_0\eqdef 0$.
The nodes labeled ``$\calN$'' represent 
zero-mean normal densities with variance~1;
the node labeled ``$\calN(0,\sigma^2 I)$''
represents a zero-mean multivariate normal density with covariance matrix $\sigma^2 I$.
All other nodes in \Fig{fig:FGCS} represent deterministic constraints.

For fixed $\sigma_1,\ldots,\sigma_K$,
\Fig{fig:FGCS} is a cycle-free linear Gaussian factor graph
and MAP/MMSE/LMMSE estimation (of any variables) 
can be carried out by Gaussian message passing,
as described in detail in \cite{LDHKLK:fgsp2007}.
Interestingly, in this particular example, most of the message 
passing can be carried out symbolically, i.e., as a technique
to derive closed-form expressions for the estimates.

Every message in this paper is a (scalar or multivariate) Gaussian distribution, 
up to a scale factor.
(Sometimes, we also allow a degenerate limit of a Gaussian,
such as a ``Gaussian'' with variance zero or infinity, 
but we will not discuss this in detail.)
Scale factors can be ignored in this paper. 
Messages can thus be parameterized by a mean vector and a covariance matrix.
For example, $\msgf{m}{X_k}$ and $\msgf{V}{X_k}$ denote
the mean vector and the covariance matrix, respectively, of the message 
traveling forward on the edge $X_k$ in \Fig{fig:FGCS}, 
while $\msgb{m}{X_k}$ and $\msgb{V}{X_k}$ denote 
the mean vector and the covariance matrix, respectively,
of the message traveling backward on the edge $X_k$.
Alternatively, messages can be parameterized by the 
precision matrix $\msgf{W}{X_k}$ 
(= the inverse of the covariance matrix $\msgf{V}{X_k}$)
and the precision-weighted mean vector 
\begin{equation}
\msgf{\xi}{X_k} \eqdef \msgf{W}{X_k} \msgf{m}{X_k}.
\end{equation}
Again, the backward message along the same edge will be denoted 
by reversed arrows.

In a directed graphical model without cycles as in \Fig{fig:FGCS},
forward messages represent priors while backward messages represent likelihood functions
(up to a scale factor).

In addition, we also work with marginals of the posterior distribution 
(i.e., the product of forward message and backward message along 
the same edge \cite{LDHKLK:fgsp2007,Lg:ifg2004}).
For example, $m_{X_k}$ and $V_{X_k}$ denote the posterior mean vector and the 
posterior covariance matrix, respectively, of $X_k$. 
An important role in this paper is played by the alternative parameterization
with the dual precision matrix
\begin{equation}
\dW_{X_k} \eqdef \big( \msgf{V}{X_k} + \msgb{V}{X_k} \big)^{-1}
\end{equation}
and the dual mean vector
\begin{equation} \label{eqn:dxiDef}
\dxi_{X_k} \eqdef \dW_{X_k} (\msgf{m}{X_k} - \msgb{m}{X_k}).
\end{equation}

Message computations with all these parameterizations 
are given in Tables \ref{tab:EquGMP}--\ref{tab:InpGMP} in Appendix~\ref{appsec:GMPTables},
which contain numerous improvements over 
the corresponding tables in \cite{LDHKLK:fgsp2007}.

\section{Linear State Space Models}
\label{sec:StateSpaceModels}

\begin{figure}
\centering
\setlength{\unitlength}{0.9mm}
\begin{picture}(65,80)(0,0)

\put(-2,42.5){\pos{cr}{$\cdots$}}
\put(0,42.5){\vector(1,0){12}}      \put(0,43.5){\pos{bl}{$X_{k-1}$}}
\put(12,39.5){\framebox(6,6){$A$}}
\put(18,42.5){\vector(1,0){12}}
\put(30,40){\framebox(5,5){$+$}}
 \put(29.5,69.5){\framebox(6,6){$\calN$}}
 \put(32.5,69.5){\vector(0,-1){9}}    \put(33.5,65){\pos{cl}{$U_k$}}
 \put(29.5,54.5){\framebox(6,6){$B$}}
 \put(32.5,54.5){\vector(0,-1){9.5}}
\put(24,35){\dashbox(17,45){}}   

\put(35,42.5){\vector(1,0){12.5}}
\put(47.5,40){\framebox(5,5){$=$}}
 \put(50,40){\vector(0,-1){9.5}}
 \put(47,24.5){\framebox(6,6){$C$}}
 \put(50,24.5){\vector(0,-1){9.5}}    \put(51,19.75){\pos{cl}{$\tilde Y_k$}}
 \put(47.5,10){\framebox(5,5){$+$}}
  \put(30,9.5){\framebox(6,6){}}      \put(33,16.5){\pos{cb}{$\calN(0,\sigma^2 I)$}}
  \put(36,12.5){\vector(1,0){11.5}}   \put(41.75,11.25){\pos{ct}{$Z_k$}}
 \put(50,10){\line(0,-1){10}}
   \put(50,10){\vector(0,-1){8}}
 \put(50,1){\knownBox}             \put(52,0){\pos{bl}{$Y_k = y_k$}}

\put(52.5,42.5){\line(1,0){12.5}}   \put(60,43.5){\pos{bc}{$X_k$}}
\put(67,42.5){\pos{cl}{$\cdots$}}

\end{picture}
\vspace{1mm}
\caption{\label{fig:StateSpaceModel}%
One section of the factor graph of the linear state space model 
(\ref{eqn:StateSpaceModelState}) and~(\ref{eqn:StateSpaceModelOutput}).
The whole factor graph consists of many such sections 
and optional initial and/or terminal conditions.
The dashed block will be varied in Section~\ref{sec:StateSpaceNUV}.}
\end{figure}
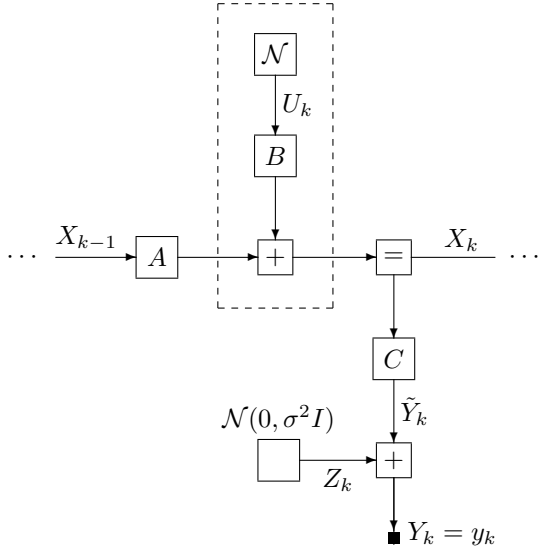

Consider a standard linear state space model 
with state $X_k \in \R^n$ and observation $Y_k\in \R^L$
evolving according to
\begin{IEEEeqnarray}{rCl}
X_k & = & A X_{k-1} + B U_k  \label{eqn:StateSpaceModelState}\\
Y_k & = & C X_k + Z_k     \label{eqn:StateSpaceModelOutput}
\end{IEEEeqnarray}
with $A\in \R^{n\times n}$, $B\in\R^{n\times m}$, $C\in\R^{L\times n}$,
and where $U_k$ (with values in $\R^m$) and $Z_k$ (with values in $\R^L$) 
are independent zero-mean white Gaussian noise processes.
We will usually assume, first, that $L=1$, 
and second, that the covariance matrix of $U_k$ is an identity matrix,
but these assumptions are not essential.
A cycle-free factor graph of such a model is shown in \Fig{fig:StateSpaceModel}.

In Section~\ref{sec:StateSpaceNUV}, 
we will vary and augment such models with NUV priors on various quantities.

Inference in such a state space model amounts to 
Kalman filtering and smoothing \cite{RoGha:unifLGM1999,KSH:LinEs2000b}
or, equivalently, to Gaussian message passing in the factor graph of \Fig{fig:StateSpaceModel} 
\cite{Lg:ifg2004,LDHKLK:fgsp2007}.
(Estimating the input $U_k$ is not usually considered 
in the Kalman filter literature, but it is essential for signal processing,
cf.\ \cite{BrLg:sise2014c,Diss:Bruderer2015}.)
With the tables in the appendix, it is easy to put together 
a large variety of such algorithms. The relative merits
of different such algorithms depend on the particulars of the problem.
However, we find the following two algorithms
usually to be the most advantageous, both in terms of computational complexity 
and in terms of numerical stability. 
If both the input $U_k$ and output $Y_k$ are scalar 
(or can be decomposed into multiple scalar inputs and outputs),
neither of these two algorithms requires a matrix inversion.
The first of these algorithms is essentially 
the Modified Bryson--Frazier (MBF) smoother \cite{Bi:fmdse}
augmented with input-signal estimation.

\noindent
\begin{trivlist}
\item{\bf MBF Message Passing:}
\begin{enumerate}
\item
Perform forward message passing with $\msgf{m}{X_k}$ and $\msgf{V}{X_k}$
using (\ref{eqn:TabAddGMPmf}), (\ref{eqn:TabAddGMPVf}), 
(\ref{eqn:TabMatMultGMPmf}), (\ref{eqn:TabMatMultGMPVf}), 
(\ref{eqn:TabObsGMPmf}), (\ref{eqn:TabObsGMPVf}).

(This is the standard Kalman filter.)
\item
Perform backward message passing with $\dxi_{X_k}$ and $\dW_{X_k}$,
beginning with $\dxi_{X_N}=0$ and $\dW_{X_N}=0$ 
at the end of the horizon, using
(\ref{eqn:TabAddGMPdxi}), (\ref{eqn:TabAddGMPdW}), 
(\ref{eqn:TabMatMultGMPdxi}), (\ref{eqn:TabMatMultGMPdW}), 
and either (\ref{eqn:TabObsGMPdxiZ}), (\ref{eqn:TabObsGMPdWZ}), (\ref{eqn:TabObsGMPF}) 
or (\ref{eqn:TabObsGMPdxi}), (\ref{eqn:TabObsGMPdW}), (\ref{eqn:TabObsGMPE}).

%
\item
Inputs $U_k$ may then be estimated 
using (\ref{eqn:TabAddGMPdxi}), (\ref{eqn:TabAddGMPdW}), 
(\ref{eqn:TabMatMultGMPdxi}), (\ref{eqn:TabMatMultGMPdW}),
(\ref{eqn:TabSingleEdgeGMPm}), (\ref{eqn:TabSingleEdgeGMPV2}).
\item
The posterior mean $m_{X_k}$ and covariance matrix $V_{X_k}$ 
of any state $X_k$ (or of an individual component thereof) may be obtained 
from  (\ref{eqn:TabSingleEdgeGMPm}) and (\ref{eqn:TabSingleEdgeGMPV2})
%
\item
Outputs \mbox{$\tilde Y_k \eqdef C X_k$} may then (very obviously) be estimated using
(\ref{eqn:TabEquGMPm}), (\ref{eqn:TabEquGMPV}), (\ref{eqn:TabMatMultGMPm}), (\ref{eqn:TabMatMultGMPV}).
\end{enumerate}
\vspace{-2ex}\hfill$\Box$%
\end{trivlist}

In step~2, the initialization with $\dW_{X_N}=0$ corresponds to the typical situation
with no \emph{a priori} information about the state $X_N$ at the end of the horizon.
MBF message passing is especially attractive for input signal estimation 
(as in step~3 above), without steps 4 and~5.

The second algorithm is an exact dual to MBF message passing and especially 
attractive for state estimation and output signal estimation 
(i.e., for standard Kalman smoothing), without steps 4 and~5 below. 
This algorithm---backward recursion with time-reversed information filter,
forward recursion with marginals (BIFM)---does not seem to be widely known. 
\begin{trivlist}
\item{\bf BIFM Message Passing:}
\begin{enumerate}
\item
Perform backward message passing with $\msgb{\xi}{X_k}$ and $\msgb{W}{X_k}$
using (\ref{eqn:TabEquGMPxib}), (\ref{eqn:TabEquGMPWb}), 
(\ref{eqn:TabMatMultGMPxib}), (\ref{eqn:TabMatMultGMPWb}), 
and (\ref{eqn:InpGMPxiZ}), (\ref{eqn:InpGMPWZ}) 
with the changes ``for the reverse direction'' stated in Table~\ref{tab:InpGMP}.
(This is a time-reversed version of the standard information filter.)
\item
Perform forward message passing with $m_{X_k}$ and $V_{X_k}$
using
(\ref{eqn:TabEquGMPm}), (\ref{eqn:TabEquGMPV}), 
(\ref{eqn:TabMatMultGMPm}), (\ref{eqn:TabMatMultGMPV}), 
and either (\ref{eqn:TabInpGMPmXZ}), (\ref{eqn:TabInpGMPVXZ}), (\ref{eqn:TabInpGMPdF}) 
or (\ref{eqn:TabInpGMPmX}), (\ref{eqn:TabInpGMPVX}), (\ref{eqn:TabInpGMPdF2}).
%
\item
Outputs $\tilde Y_k$ may then (very obviously) be estimated using
(\ref{eqn:TabEquGMPm}), (\ref{eqn:TabEquGMPV}), (\ref{eqn:TabMatMultGMPm}), (\ref{eqn:TabMatMultGMPV}).
\item
The dual means $\dxi_{X_k}$ and the dual precision matrices $\dW_{X_k}$ may be obtained 
from (\ref{eqn:TabSingleEdgeGMPdxi2}) and (\ref{eqn:TabSingleEdgeGMPdW2}).
\item
Inputs $U_k$ may then be estimated 
using (\ref{eqn:TabAddGMPdxi}), (\ref{eqn:TabAddGMPdW}), 
(\ref{eqn:TabMatMultGMPdxi}), (\ref{eqn:TabMatMultGMPdW}),
(\ref{eqn:TabSingleEdgeGMPm}), (\ref{eqn:TabSingleEdgeGMPV2}).
\end{enumerate}
\vspace{-2ex}\hfill$\Box$%
\end{trivlist}

\section{Sparsity by NUV in State Space Models}
\label{sec:StateSpaceNUV}

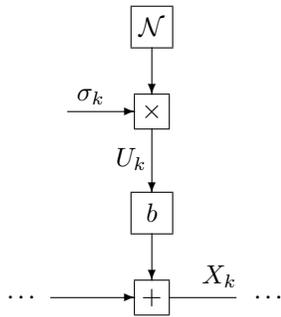
\begin{figure}[p]
\centering
\setlength{\unitlength}{0.9mm}
\begin{picture}(27.5,45.5)(0,0)
%
\put(12,39.5){\framebox(6,6){$\calN$}}
 \put(15,39.5){\vector(0,-1){7}}
\put(12.5,27.5){\framebox(5,5){$\times$}}
 \put(2.5,30){\vector(1,0){10}}     \put(6,31){\pos{bc}{$\sigma_{k}$}}
\put(15,27.5){\vector(0,-1){9.5}}   \put(14.25,23){\pos{cr}{$U_{k}$}}
\put(12,12){\framebox(6,6){$b$}}
\put(15,12){\vector(0,-1){7}}

\put(-4,2.5){\cent{\ldots}}
\put(0,2.5){\vector(1,0){12.5}} 
\put(12.5,0){\framebox(5,5){$+$}}
\put(17.5,2.5){\line(1,0){10}}      \put(25,4){\pos{cb}{$X_k$}}
\put(32.5,2.5){\cent{\ldots}}
\end{picture}
\vspace{1mm}
\caption{\label{fig:SparseInputSignal}%
Alternative input block (to replace the dashed box in \Fig{fig:StateSpaceModel}) 
for a sparse scalar input signal $U_1,U_2,\ldots$.}
\end{figure}

Sparse input signals are easily introduced: 
simply replace the normal prior on $U_k$ 
in (\ref{eqn:StateSpaceModelState}) and in \Fig{fig:StateSpaceModel} by a NUV prior, 
as shown in \Fig{fig:SparseInputSignal}. 
This approach was used in \cite{BrMaLg:ISIT2015c} 
to estimate the input signal $U_1,U_2,\ldots$ itself.

However, we may also be interested in the clean output signal 
$\tilde Y_k = C X_k$.
For example, consider the problem of 
approximating some given signal $y_1,y_2,\ldots \in \R$ 
by constant segments, as illustrated in \Fig{fig:LevelJumps}.
The constant segments can be represented 
by the simplest possible state space model with $n=1$, $A = C = (1)$,
and no input. For the occasional jumps between the constant segments,
we use a sparse input signal $U_1,U_2,\ldots$ with a NUV prior
(and with $B=b=(1)$) as in \Fig{fig:SparseInputSignal}.
The sparsity level---i.e., the number of constant segments---%
can be controlled by the assumed observation noise $\sigma^2$.

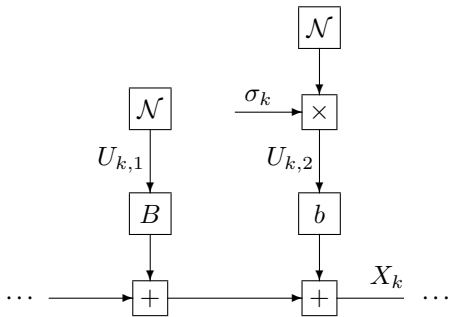
\begin{figure}
\centering
\setlength{\unitlength}{0.9mm}
\begin{picture}(52.5,45.5)(0,0)
%
\put(12,27.5){\framebox(6,6){$\calN$}}
\put(15,27.5){\vector(0,-1){9.5}}   \put(14.25,23){\pos{cr}{$U_{k,1}$}}
\put(12,12){\framebox(6,6){$B$}}
\put(15,12){\vector(0,-1){7}}

\put(37,39.5){\framebox(6,6){$\calN$}}
 \put(40,39.5){\vector(0,-1){7}}
\put(37.5,27.5){\framebox(5,5){$\times$}}
 \put(27.5,30){\vector(1,0){10}}    \put(31,31){\pos{bc}{$\sigma_{k}$}}
\put(40,27.5){\vector(0,-1){9.5}}    \put(39.25,23){\pos{cr}{$U_{k,2}$}}
\put(37,12){\framebox(6,6){$b$}}
\put(40,12){\vector(0,-1){7}}

\put(-4,2.5){\cent{\ldots}}
\put(0,2.5){\vector(1,0){12.5}} 
\put(12.5,0){\framebox(5,5){$+$}}
\put(17.5,2.5){\vector(1,0){20}}
\put(37.5,0){\framebox(5,5){$+$}}
\put(42.5,2.5){\line(1,0){10}}      \put(50,4){\pos{cb}{$X_k$}}
\put(57.5,2.5){\cent{\ldots}}
\end{picture}
\vspace{1mm}
\caption{\label{fig:RandomWalkInputModel}%
Input block with both white noise and additional sparse scalar input.}
\end{figure}

The sparse scalar input signal of \Fig{fig:SparseInputSignal}
can be generalized in several different directions. 
A first obvious generalization is to combine a primary white-noise 
input with a secondary sparse input as shown in \Fig{fig:RandomWalkInputModel}.
For example, the constant segments in \Fig{fig:LevelJumps}
are thus generalized to random-walk segments as in \Fig{fig:WalkJumps}.

\begin{figure}
\centering
\setlength{\unitlength}{0.9mm}
\begin{picture}(52.5,45.5)(0,0)
%
\put(12,39.5){\framebox(6,6){$\calN$}}
 \put(15,39.5){\vector(0,-1){7}}
\put(12.5,27.5){\framebox(5,5){$\times$}}
 \put(2.5,30){\vector(1,0){10}}     \put(6,31){\pos{bc}{$\sigma_{k,1}$}}
\put(15,27.5){\vector(0,-1){9.5}}   \put(14.25,23){\pos{cr}{$U_{k,1}$}}
\put(12,12){\framebox(6,6){$b_1$}}
\put(15,12){\vector(0,-1){7}}

\put(37,39.5){\framebox(6,6){$\calN$}}
 \put(40,39.5){\vector(0,-1){7}}
\put(37.5,27.5){\framebox(5,5){$\times$}}
 \put(27.5,30){\vector(1,0){10}}    \put(31,31){\pos{bc}{$\sigma_{k,2}$}}
\put(40,27.5){\vector(0,-1){9.5}}    \put(39.25,23){\pos{cr}{$U_{k,2}$}}
\put(37,12){\framebox(6,6){$b_2$}}
\put(40,12){\vector(0,-1){7}}

\put(-4,2.5){\cent{\ldots}}
\put(0,2.5){\vector(1,0){12.5}} 
\put(12.5,0){\framebox(5,5){$+$}}
\put(17.5,2.5){\vector(1,0){20}}
\put(37.5,0){\framebox(5,5){$+$}}
\put(42.5,2.5){\line(1,0){10}}      \put(50,4){\pos{cb}{$X_k$}}
\put(57.5,2.5){\cent{\ldots}}
\end{picture}
\vspace{1mm}
\caption{\label{fig:LinesInputModel}%
Input block with two separate sparse scalar inputs 
for two degrees of freedom such as in \Fig{fig:LinesJumps}.}
\end{figure}
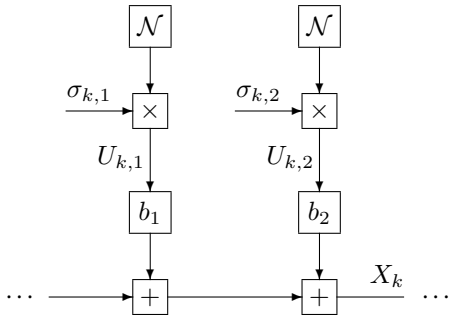

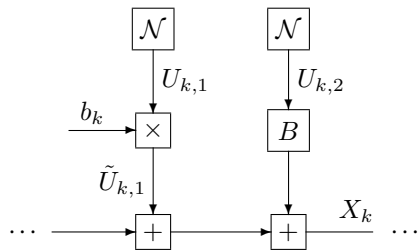
\begin{figure}
\centering
\setlength{\unitlength}{0.9mm}
\begin{picture}(47.5,35.5)(0,0)

\put(12,29.5){\framebox(6,6){$\calN$}}
\put(15,29.5){\vector(0,-1){9.5}}       \put(16.25,24.75){\pos{cl}{$U_{k,1}$}}
 \put(2.5,17.5){\vector(1,0){10}}       \put(6,18.5){\pos{bc}{$b_k$}}
\put(12.5,15){\framebox(5,5){$\times$}}
\put(15,15){\vector(0,-1){10}}          \put(14,10){\pos{cr}{$\tilde U_{k,1}$}}

\put(32,29.5){\framebox(6,6){$\calN$}}
\put(35,29.5){\vector(0,-1){9}}         \put(36.25,24.75){\pos{cl}{$U_{k,2}$}}
\put(32,14.5){\framebox(6,6){$B$}}
\put(35,14.5){\vector(0,-1){9.5}}

\put(-4,2.5){\cent{\ldots}}
\put(0,2.5){\vector(1,0){12.5}}
\put(12.5,0){\framebox(5,5){$+$}}
\put(17.5,2.5){\vector(1,0){15}}
\put(32.5,0){\framebox(5,5){$+$}}
\put(37.5,2.5){\line(1,0){10}}    \put(45,4){\pos{cb}{$X_k$}}
\put(52.5,2.5){\cent{\ldots}}

\end{picture}
\vspace{1mm}
\caption{\label{fig:EventB}%
Input block allowing general sparse pulses, each with its own 
signature~$b_k$, in addition to full-rank white noise.}
\end{figure}


\begin{figure}
\centering
\setlength{\unitlength}{0.9mm}
\begin{picture}(50,57.5)(15,0)

\put(13.5,55){\cent{\ldots}}
\put(17.5,55){\vector(1,0){12.5}}   \put(20,53){\pos{ct}{$X_k$}}
\put(30,52.5){\framebox(5,5){$=$}}
\put(35,55){\line(1,0){15}}
\put(55,55){\cent{\ldots}}

\put(32.5,52.5){\vector(0,-1){9.5}}
\put(29.5,37){\framebox(6,6){$C$}}
\put(32.5,37){\vector(0,-1){9.5}}       \put(33.5,32.25){\pos{cl}{$\tilde Y_k$}}
\put(30,22.5){\framebox(5,5){$+$}}
\put(32.5,22.5){\vector(0,-1){7.5}}

\put(14.5,22){\framebox(6,6){$\calN$}}
\put(20.5,25){\vector(1,0){9.5}}        \put(25.25,23.75){\pos{ct}{$Z_k$}}

\put(59,9.5){\framebox(6,6){$\calN$}}
\put(59,12.5){\vector(-1,0){9}}
\put(45,10){\framebox(5,5){$\times$}}
 \put(47.5,27.5){\vector(0,-1){12.5}}    \put(49,25){\pos{cl}{$\sigma_k$}}
\put(45,12.5){\vector(-1,0){10}}         \put(40,13.75){\pos{cb}{$\tilde Z_k$}}

\put(30,10){\framebox(5,5){$+$}}
\put(32.5,10){\line(0,-1){10}}
 \put(32.5,10){\vector(0,-1){8}}
\put(32.5,1){\knownBox}             \put(35,1){\pos{cl}{$y_k$}}
\end{picture}
\vspace{1mm}
\caption{\label{fig:Outliers}%
Alternative output block for scalar signal with outliers.}
\end{figure}
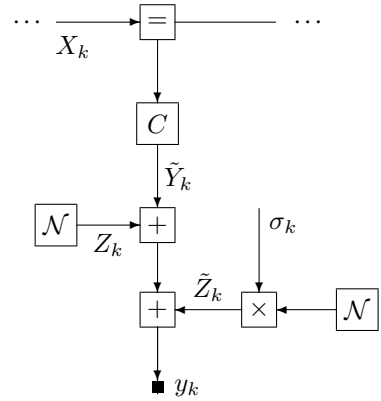

\begin{figure}
\centering
\includegraphics[width=0.8\linewidth]{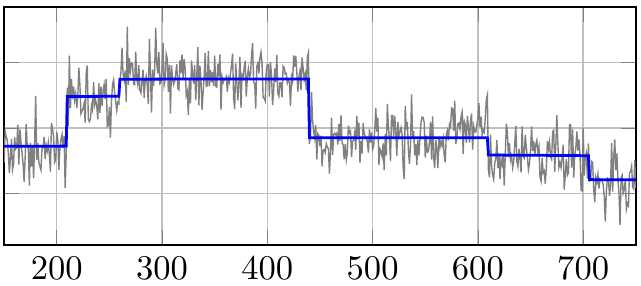}
\caption{\label{fig:LevelJumps}%
Estimating (or fitting) a piecewise constant signal.}
%
\vspace{5mm}
%
\includegraphics[width=0.8\linewidth]{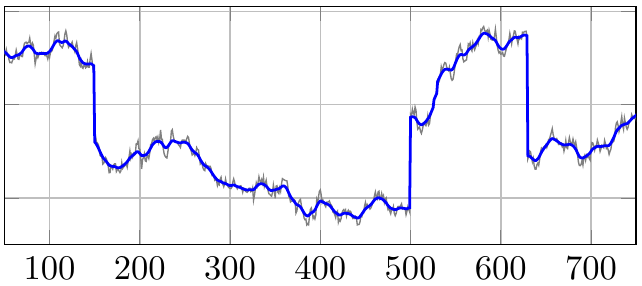}
\caption{\label{fig:WalkJumps}%
Estimating a random walk with occasional jumps.}
%
\vspace{5mm}
%
\includegraphics[width=0.8\linewidth]{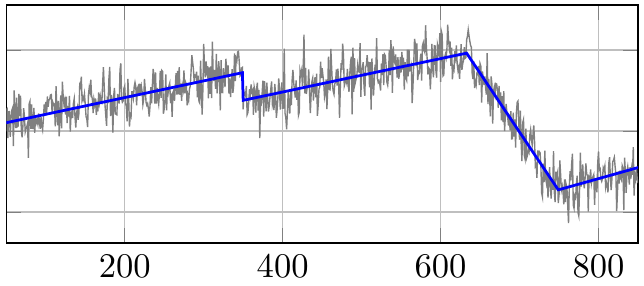}
\caption{\label{fig:LinesJumps}%
Approximation with straight-line segments.}
\vspace{5mm}
%
\includegraphics[width=0.8\linewidth]{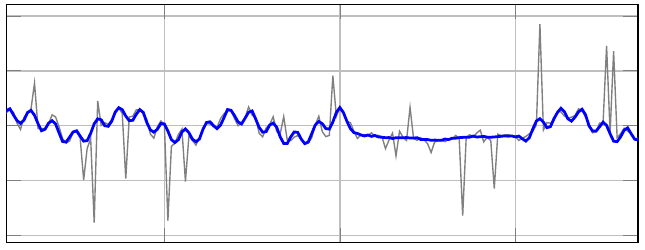}
\caption{\label{fig:OutliersPlot}%
Outlier removal according to \Fig{fig:Outliers}.}
\end{figure}

Another generalization of \Fig{fig:LevelJumps} 
is shown in \Fig{fig:LinesJumps}, where the constant-level segments 
are replaced by straight-line segments, which can be represented 
by a state space model of order $n=2$. 
The corresponding input block, 
with two separate sparse scalar input 
signals, is shown in \Fig{fig:LinesInputModel};
the first input, $U_{k,1},$ affects the magnitude 
and the second input, $U_{k,2},$ affects the slope of the line model.
The further generalization to polynomial segments is obvious. 
Continuity can be enforced by omitting the input $U_{k,1},$
and continuity of derivatives can be enforced likewise.

More generally, \Fig{fig:LinesInputModel} 
(generalized to an arbitrary number of sparse scalar input signals)
can be used to allow occasional jumps in individual components of the state
of arbitrary state space models.

In all these examples, the parameters $\sigma_{k}^2$ 
(or $\sigma_{k,\ell}^2$)
can be learned 
as described in Section~\ref{sec:VarEst},
and the required quantities $m_{U_k}$ and $\sigma_{U_k}^2$
(or $m_{U_{k,\ell}}$ and $\sigma_{U_{k,\ell}}^2,$ respectively)
can be computed by message passing in the pertinent factor graph 
as described in Section~\ref{sec:StateSpaceModels}.

A more substantial generalization of \Fig{fig:SparseInputSignal}
is shown in \Fig{fig:EventB}, with $\sigma_k$ of \Fig{fig:SparseInputSignal}
generalized to $b_k\in \R^n$. 
We mention without proof that this generalized NUV prior 
on $\tilde U_{k,1} \eqdef b_k U_{k,1}$ still promotes sparsity
and can be learned by EM
(provided that $BB^\T$ has full rank)
\cite{ZML:spICASSP2016c}.
This input model allows quite general events to happen,
each with its own signature $b_k$. 
The estimated nonzero vectors $\hat b_1, \hat b_2,\ldots$ may be viewed 
as features of the given signal $y_1,y_2,\ldots$ 
that can be used for further analysis.

Finally, we turn to the output block in \Fig{fig:StateSpaceModel}.
A simple and effective method to detect and to remove outliers 
from the scalar output signal of a state space model 
is to replace (\ref{eqn:StateSpaceModelOutput}) with
\begin{equation}
Y_k = C X_k + Z_k + \tilde Z_k
\end{equation}
with sparse $\tilde Z_k$, 
as shown in \Fig{fig:Outliers} \cite{WaLg:OIKSd2016}. 
Again, the parameters $\sigma_k$ can be estimated 
by EM essentially as described in Section~\ref{sec:VarEst},
and the required quantities $m_{\tilde Z_k}$ and $\sigma^2_{\tilde Z_k}$
can be computed by message passing as described in Section~\ref{sec:StateSpaceModels}.
An example of this method is shown in \Fig{fig:OutliersPlot}
for some state space model of order $n=4$ 
with details that are irrelevant for this paper.

\section{Conclusion}

We have given improved tables of Gaussian-message computations 
for estimation in linear state space models, 
and we have pointed out 
two preferred message passing algorithms: 
the first algorithm is essentially the Modified Bryson-Frazier smoother,
the second algorithm is a dual of it.
In addition, we have advocated NUV priors (together with EM algorithms) 
from sparse Bayesian learning 
for introducing sparsity into linear state space models
and outlined several applications.

In this paper, all factor graphs were cycle-free 
so that Gaussian message passing yields exact marginals. 
The use of NUV regularization in factor graphs with cycles,
and its relative merits in comparison with, e.g., AMP \cite{DMM:mpcs},
remains to be investigated.

\appendices
\section{Tabulated Gaussian-Message Computations}
\label{appsec:GMPTables}


Tables~\ref{tab:EquGMP}--\ref{tab:InpGMP} 
are improved versions of the corresponding tables in \cite{LDHKLK:fgsp2007}.
The notation for the different parameterizations of the messages 
was defined in Section~\ref{sec:FG}. 
The main novelties of this new version are the following:

\begin{enumerate}
\item
New notation
$\msgf{\xi}{} \eqdef \msgf{W}{} \msgf{m}{}$ and $\msgb{\xi}{} \eqdef \msgb{W}{} \msgb{m}{}$.
\item
Introduction of the dual marginal $\dxi$ (\ref{eqn:TabSingleEdgeGMPdxiDef})
with pertinent new expressions in Tables \ref{tab:EquGMP}--\ref{tab:ObsGMP},
and new expressions with the dual precision matrix, 
especially (\ref{eqn:TabObsGMPdxiZ})--(\ref{eqn:TabObsGMPE}).
These results (from \cite{Diss:Bruderer2015}) are used 
both in Appendix~\ref{appsec:LeastSquaresMessagePassing}
and in the two preferred algorithms in Section~\ref{sec:StateSpaceModels}.
\item
New expressions (\ref{eqn:TabInpGMPmXZ})--(\ref{eqn:TabInpGMPdF2}) 
for the marginals, which are essential for the BIFM Kalman smoother in Section~\ref{sec:StateSpaceModels}.
\end{enumerate}

\begin{table}[t]
\Beginlocaleqncntr
\caption{\label{tab:EquGMP}%
Gaussian message passing through an equality-constraint.}
\normalsize%
\framebox[\linewidth]{%
\begin{minipage}{0.95\linewidth}
\begin{center}
\setlength{\unitlength}{0.9mm}
\begin{picture}(30,20)(0,0)
%
\put(0,12.5){\vector(1,0){12.5}}      \put(5,14){\pos{bc}{$X$}}
\put(12.5,10){\framebox(5,5){$=$}}
 \put(15,10){\vector(0,-1){10}}       \put(14,2){\pos{br}{$Y$}}
\put(17.5,12.5){\vector(1,0){12.5}}   \put(25,14){\pos{bc}{$Z$}}
\end{picture}
\end{center}
Constraint $X=Y=Z$,\\
expressed by factor $\delta(z-x)\,\delta(y-x)$
\vspace{1ex}

{\gray\hrule\vspace{-1.5ex}}
\begin{IEEEeqnarray}{rCl}
\rule{0em}{4ex}%
\msgf{\xi}{Z} & = & \msgf{\xi}{X} + \msgb{\xi}{Y} \label{eqn:TabEquGMPxi}\\
\msgf{W}{Z} & = & \msgf{W}{X} + \msgb{W}{Y}  \label{eqn:TabEquGMPW}
\end{IEEEeqnarray}

{\gray\hrule\vspace{-1.5ex}}
\begin{IEEEeqnarray}{rCl}
\rule{0em}{4ex}%
\msgb{\xi}{X} & = & \msgb{\xi}{Y} + \msgb{\xi}{Z} \label{eqn:TabEquGMPxib}\\
\msgb{W}{X} & = & \msgb{W}{Y} + \msgb{W}{Z}  \label{eqn:TabEquGMPWb}
\end{IEEEeqnarray}

{\gray\hrule\vspace{-1.5ex}}
\begin{IEEEeqnarray}{rCcCl}
m_X & = & m_Y & = & m_Z  \label{eqn:TabEquGMPm}\\
V_X & = & V_Y & = & V_Z  \label{eqn:TabEquGMPV}
\end{IEEEeqnarray}

{\gray\hrule\vspace{-1.5ex}}
\begin{IEEEeqnarray}{rCl}
\rule{0em}{3.5ex}%
\dxi_X & = & \dxi_Y + \dxi_Z \label{eqn:TabEquGMPdxi}
\end{IEEEeqnarray}
\vspace{-3.5ex}
\end{minipage}%
}
\Endlocaleqncntr
\end{table}

\begin{table}[t]
\Beginlocaleqncntr
\caption{\label{tab:AddGMP}%
Gaussian message passing through an adder node.}
\normalsize%
\framebox[\linewidth]{%
\begin{minipage}{0.95\linewidth}
\begin{center}
\setlength{\unitlength}{0.9mm}
\begin{picture}(30,20)(0,0)
%
\put(0,12.5){\vector(1,0){12.5}}      \put(5,14){\pos{bc}{$X$}}
\put(12.5,10){\framebox(5,5){$+$}}
 \put(15,0){\vector(0,1){10}}       \put(14,2){\pos{br}{$Y$}}
\put(17.5,12.5){\vector(1,0){12.5}}   \put(25,14){\pos{bc}{$Z$}}
\end{picture}
\end{center}
Constraint $Z=X+Y$,\\ 
expressed by factor $\delta(z-(x+y))$
\vspace{1ex}

{\gray\hrule\vspace{-1.5ex}}
\begin{IEEEeqnarray}{rCl}
\rule{0em}{3.5ex}%
\msgf{m}{Z} & = & \msgf{m}{X} + \msgf{m}{Y}  \label{eqn:TabAddGMPmf}\\
\msgf{V}{Z} & = & \msgf{V}{X} + \msgf{V}{Y}  \label{eqn:TabAddGMPVf}
\end{IEEEeqnarray}

{\gray\hrule\vspace{-1.5ex}}
\begin{IEEEeqnarray}{rCl}
\rule{0em}{3.5ex}%
\msgb{m}{X} & = & \msgb{m}{Z} - \msgf{m}{Y} \label{eqn:TabAddGMPmb}\\
\msgb{V}{X} & = & \msgb{V}{Z} + \msgf{V}{Y} \label{eqn:TabAddGMPVb}
\end{IEEEeqnarray}

{\gray\hrule\vspace{-1.5ex}}
\begin{IEEEeqnarray}{rCl}
m_Z & = & m_X + m_Y
\end{IEEEeqnarray}

{\gray\hrule\vspace{-1.5ex}}
\begin{IEEEeqnarray}{rCcCl}
\rule{0em}{3.5ex}%
\dxi_X & = & \dxi_Y & = & \dxi_Z  \label{eqn:TabAddGMPdxi}\\
\dW_X & = & \dW_Y & = & \dW_Z  \label{eqn:TabAddGMPdW}
\end{IEEEeqnarray}
\vspace{-3.5ex}
\end{minipage}%
}
\Endlocaleqncntr
\end{table}

\begin{table}[t]
\Beginlocaleqncntr
\caption{\label{tab:MMultGMP}%
Gaussian message passing through a matrix multiplier node with arbitrary real matrix $A$.}
\normalsize%
\framebox[\linewidth]{%
\begin{minipage}{0.95\linewidth}
\begin{center}
\setlength{\unitlength}{0.9mm}
\begin{picture}(30,10.5)(0,0)
%
\put(0,3){\vector(1,0){12}}      \put(5,4.5){\pos{bc}{$X$}}
\put(12,0){\framebox(6,6){$A$}}
\put(18,3){\vector(1,0){12}}     \put(25,4.5){\pos{bc}{$Y$}}
\end{picture}
\vspace{3.5mm}

Constraint $Y=AX$,
expressed by factor $\delta(y-Ax)$
\end{center}

{\gray\hrule\vspace{-1.5ex}}
\begin{IEEEeqnarray}{rCl}
\rule{0em}{3.5ex}%
\msgf{m}{Y} & = & A \msgf{m}{X}   \label{eqn:TabMatMultGMPmf}\\
\msgf{V}{Y} & = & A \msgf{V}{X} A^\T    \label{eqn:TabMatMultGMPVf}
\end{IEEEeqnarray}

{\gray\hrule\vspace{-1.5ex}}
\begin{IEEEeqnarray}{rCl}
\rule{0em}{3.5ex}%
\msgb{\xi}{X} & = & A^\T \msgb{\xi}{Y} \label{eqn:TabMatMultGMPxib}\\
\msgb{W}{X} & = & A^\T \msgb{W}{Y} A   \label{eqn:TabMatMultGMPWb}
\end{IEEEeqnarray}

{\gray\hrule\vspace{-1.5ex}}
\begin{IEEEeqnarray}{rCl}
m_Y & = & A m_X        \label{eqn:TabMatMultGMPm}\\
V_Y & = & A V_X A^\T   \label{eqn:TabMatMultGMPV}
\end{IEEEeqnarray}

{\gray\hrule\vspace{-1.5ex}}
\begin{IEEEeqnarray}{rCl}
\rule{0em}{3.5ex}%
\dxi_X & = & A^\T \dxi_Y \label{eqn:TabMatMultGMPdxi}\\
\dW_X & = & A^\T \dW_Y A  \label{eqn:TabMatMultGMPdW}
\end{IEEEeqnarray}
\vspace{-3.5ex}
\end{minipage}%
}
\Endlocaleqncntr
\end{table}

\begin{table}[t]
\Beginlocaleqncntr
\caption{\label{tab:SingleEdgeGMP}%
Gaussian single-edge marginals ($m$, $V$) and their duals ($\dxi$, $\dW$).}
\normalsize%
\framebox[\linewidth]{%
\begin{minipage}{0.95\linewidth}

\vspace{-1.5ex}
\begin{IEEEeqnarray}{rCl}
\dxi_X & \eqdef & \dW_X (\msgf{m}{X} - \msgb{m}{X})\label{eqn:TabSingleEdgeGMPdxiDef} \\
       & = & \msgf{\xi}{X} - \msgf{W}{X} m_X \label{eqn:TabSingleEdgeGMPdxi}\\
       & = &  \msgb{W}{X} m_X - \msgb{\xi}{X} \label{eqn:TabSingleEdgeGMPdxi2}\\
\dW_X & \eqdef & (\msgf{V}{X} + \msgb{V}{X})^{-1} \label{eqn:TabSingleEdgeGMPdWDef}\\
  & = & \msgf{W}{X} V_X \msgb{W}{X} \\
  & = & \msgf{W}{X} - \msgf{W}{X} V_X \msgf{W}{X} \label{eqn:TabSingleEdgeGMPdW} \\
  & = & \msgb{W}{X} - \msgb{W}{X} V_X \msgb{W}{X} \label{eqn:TabSingleEdgeGMPdW2}
  \IEEEeqnarraynumspace
\end{IEEEeqnarray}

{\gray\hrule\vspace{-1.5ex}}
\begin{IEEEeqnarray}{rCl}
\rule{0em}{3.5ex}%
m_X & = & V_X (\msgf{\xi}{X} + \msgb{\xi}{X}) \\
    & = & \msgf{m}{X} - \msgf{V}{X} \dxi_X  \label{eqn:TabSingleEdgeGMPm}\\
    & = & \msgb{m}{X} + \msgb{V}{X} \dxi_X \label{eqn:TabSingleEdgeGMPm2}\\
V_X & = & (\msgf{W}{X} + \msgb{W}{X})^{-1} \\
  & = & \msgf{V}{X} \dW_X \msgb{V}{X} \label{eqn:TabSingleEdgeGMPV1}\\
  & = & \msgf{V}{X} - \msgf{V}{X} \dW_X \msgf{V}{X} \label{eqn:TabSingleEdgeGMPV2} \\
  & = & \msgb{V}{X} - \msgb{V}{X} \dW_X \msgb{V}{X}
  \IEEEeqnarraynumspace
\end{IEEEeqnarray}
\vspace{-3.5ex}
\end{minipage}%
}
\Endlocaleqncntr
\end{table}

\begin{table}[t]
\Beginlocaleqncntr
\caption{\label{tab:ObsGMP}%
Gaussian message passing through an observation block.}
\normalsize%
\framebox[\linewidth]{%
\begin{minipage}{0.95\linewidth}
\begin{center}
\setlength{\unitlength}{0.9mm}
\begin{picture}(30,35)(0,0)
%
\put(0,27.5){\vector(1,0){12.5}}      \put(5,29){\pos{bc}{$X$}}
\put(12.5,25){\framebox(5,5){$=$}}
\put(17.5,27.5){\vector(1,0){12.5}}   \put(25,29){\pos{bc}{$Z$}}
\put(15,25){\vector(0,-1){8}}
\put(12,11){\framebox(6,6){$A$}}
\put(15,11){\vector(0,-1){10}}      \put(14,3){\pos{br}{$Y$}}
\end{picture}
\end{center}

{\gray\hrule\vspace{-1.5ex}}
\begin{IEEEeqnarray}{rCl}
\rule{0em}{3.5ex}%
\msgf{m}{Z} & = & \msgf{m}{X} + \msgf{V}{X} A^\T G \, (\msgb{m}{Y} - A \msgf{m}{X}) 
   \IEEEeqnarraynumspace\label{eqn:TabObsGMPmf}\\
\msgf{V}{Z} & = & \msgf{V}{X} - \msgf{V}{X} A^\T GA \msgf{V}{X} \label{eqn:TabObsGMPVZ}
      \label{eqn:TabObsGMPVf}\\
\text{with~} G & \eqdef & \big( \msgb{V}{Y} + A \msgf{V}{X} A^\T \big)^{-1}
\end{IEEEeqnarray}

{\gray\hrule\vspace{-1.5ex}}
\begin{IEEEeqnarray}{rCl}
\rule{0em}{3.5ex}%
\dxi_X & = & F^\T \dxi_Z + A^\T \msgb{W}{Y} \big( A \msgf{m}{Z} - \msgb{m}{Y} \big)
      \label{eqn:TabObsGMPdxiZ}\IEEEeqnarraynumspace\\
  & = & F^\T \dxi_Z + A^\T G \, (A \msgf{m}{X} - \msgb{m}{Y})
   \label{eqn:TabObsGMPdxi}\IEEEeqnarraynumspace\\
\dW_X & = & F^\T \dW_Z F + A^\T\msgb{W}{Y}A F
                 \IEEEeqnarraynumspace\label{eqn:TabObsGMPdWZ}\\
      & = & F^\T \dW_Z F + A^\T GA \label{eqn:TabObsGMPdW}\\
\text{with~} F & \eqdef & I- \msgf{V}{Z} A^\T\msgb{W}{Y}A \label{eqn:TabObsGMPF}\\
     & = & I - \msgf{V}{X} A^\T GA  \label{eqn:TabObsGMPE}
\end{IEEEeqnarray}

{\gray\hrule}
\rule{0em}{3ex}For the reverse direction, 
replace \rule{0em}{3ex}$\msgf{m}{Z}$ by $\msgb{m}{X}$,
$\msgf{V}{Z}$ by $\msgb{V}{X}$,
$\msgf{m}{X}$ by $\msgb{m}{Z}$,
$\msgf{V}{X}$ by $\msgb{V}{Z}$,
exchange $\dxi_X$ and $\dxi_Z$, 
exchange $\dW_X$ and $\dW_Z$, 
and change ``$+$'' to ``$-$'' in (\ref{eqn:TabObsGMPdxiZ}) and (\ref{eqn:TabObsGMPdxi}).
\end{minipage}%
}
\Endlocaleqncntr
\end{table}

\begin{table}[t]
\Beginlocaleqncntr
\caption{\label{tab:InpGMP}%
Gaussian message passing through an input block.}
\normalsize%
\framebox[\linewidth]{%
\begin{minipage}{0.95\linewidth}
\begin{center}
\setlength{\unitlength}{0.9mm}
\begin{picture}(30,35)(0,0)
%
\put(0,27.5){\vector(1,0){12.5}}      \put(5,29){\pos{bc}{$X$}}
\put(12.5,25){\framebox(5,5){$+$}}
\put(17.5,27.5){\vector(1,0){12.5}}   \put(25,29){\pos{bc}{$Z$}}
\put(15,17){\vector(0,1){8}}
\put(12,11){\framebox(6,6){$A$}}
\put(15,1){\vector(0,1){10}}        \put(14,3){\pos{br}{$Y$}}
\end{picture}
\end{center}

{\gray\hrule\vspace{-1.5ex}}
\begin{IEEEeqnarray}{rCl}
\rule{0em}{3.5ex}%
\msgf{\xi}{Z} & = & \msgf{\xi}{X} + \msgf{W}{X} AH (\msgf{\xi}{Y} - A^\T \msgf{\xi}{X})
   \IEEEeqnarraynumspace\label{eqn:InpGMPxiZ}\\
\msgf{W}{Z} & = & \msgf{W}{X} - \msgf{W}{X} AHA^\T \msgf{W}{X} 
         \IEEEeqnarraynumspace\label{eqn:InpGMPWZ}\\
\text{with~} H & \eqdef & \big( \msgf{W}{Y} + A^\T \msgf{W}{X} A \big)^{-1}
\end{IEEEeqnarray}

{\gray\hrule\vspace{-1.5ex}}
\begin{IEEEeqnarray}{rCl}
\rule{0em}{3.5ex}%
m_X & = & \dF^\T m_Z + A \msgf{V}{Y} \big( A^\T \msgf{\xi}{Z} - \msgf{\xi_Y}{} \big) 
           \label{eqn:TabInpGMPmXZ}\\
    & = & \dF^\T m_Z + AH \big( A^\T \msgf{\xi}{X} - \msgf{\xi}{Y} \big)
   \IEEEeqnarraynumspace\label{eqn:TabInpGMPmX}\\
V_X & = & \dF^\T V_Z \dF + A\msgf{V}{Y} A^\T \dF \label{eqn:TabInpGMPVXZ}\\
    & = & \dF^\T V_Z \dF + AHA^\T
       \IEEEeqnarraynumspace\label{eqn:TabInpGMPVX}\\
\text{with~} \dF & \eqdef & I - \msgf{W}{Z} A \msgf{V}{Y} A^\T \label{eqn:TabInpGMPdF}\\
    & = & I - \msgf{W}{X}AHA^\T   \label{eqn:TabInpGMPdF2}
\end{IEEEeqnarray}

{\gray\hrule}
For the reverse direction, 
replace \rule{0em}{3ex}$\msgf{\xi}{Z}$ by $\msgb{\xi}{X}$,
$\msgf{W}{Z}$ by $\msgb{W}{X}$,
$\msgf{\xi}{X}$ by $\msgb{\xi}{Z}$,
$\msgf{W}{X}$ by $\msgb{W}{Z}$,
exchange $m_X$ and $m_Z$, 
exchange $V_X$ and $V_Z$, 
and replace \rule{0em}{3ex}$\msgf{\xi}{Y}$ by $-\msgf{\xi}{Y}$.
\end{minipage}%
}
\Endlocaleqncntr
\end{table}


The proofs (below) are given only for the new expressions;
for the other proofs, we refer to \cite{LDHKLK:fgsp2007}.

\begin{proofof}{of (\ref{eqn:TabEquGMPdxi})}
Using (\ref{eqn:TabSingleEdgeGMPdxi2}), (\ref{eqn:TabEquGMPxib}), (\ref{eqn:TabEquGMPWb}), 
and (\ref{eqn:TabEquGMPm}), 
we have
\begin{IEEEeqnarray}{rCl}
\dxi_X & = & \msgb{W}{X} m_X - \msgb{\xi}{X} \\
 & = & \big( \msgb{W}{Y} + \msgb{W}{Z} \big) m_X - \big( \msgb{\xi}{Y} + \msgb{\xi}{Z} \big) \\
 & = & \big( \msgb{W}{Y} m_Y - \msgb{\xi}{Y} \big) + \big( \msgb{W}{Z} m_Z - \msgb{\xi}{Z} \big)
       \IEEEeqnarraynumspace\\
 & = &  \dxi_Y + \dxi_Z.
\vspace{-1.5ex}
\end{IEEEeqnarray}
\end{proofof}

\begin{proofof}{of (\ref{eqn:TabAddGMPdxi})}
We first note
\begin{IEEEeqnarray}{rCl}
\msgf{m}{X} - \msgb{m}{X}
 & = & \msgf{m}{X} + \msgf{m}{Y} - \msgb{m}{Z} \\
 & = & \msgf{m}{Z} - \msgb{m}{Z},
\end{IEEEeqnarray}
and (\ref{eqn:TabAddGMPdxi}) follows from (\ref{eqn:TabAddGMPdW}).
\end{proofof}

\begin{proofof}{of (\ref{eqn:TabMatMultGMPdxi})}
Using \cite[eq.~(III.9)]{LDHKLK:fgsp2007}, we have
\begin{IEEEeqnarray}{rCl}
\dxi_X & = & \dW_X (\msgf{m}{X} - \msgb{m}{X}) \\
 & = & \dW_X \msgf{m}{X} - \dW_X \msgb{m}{X} \\
 & = & A^\T \dW_Y A \msgf{m}{X} - A^\T \dW_Y \msgb{m}{Y} 
       \IEEEeqnarraynumspace\\
 & = & A^\T \dW_Y ( \msgf{m}{Y} - \msgb{m}{Y} ).
\end{IEEEeqnarray}
\eproofnegspace
\end{proofof}

\begin{proofof}{of (\ref{eqn:TabSingleEdgeGMPm}) and (\ref{eqn:TabSingleEdgeGMPdxi})}
Using (\ref{eqn:TabSingleEdgeGMPV2}) and (\ref{eqn:TabSingleEdgeGMPV1}),
we have
\begin{IEEEeqnarray}{rCl}
m_X 
 & = & V_X \msgf{\xi}{X} + V_X \msgb{\xi}{X} \\
 & = & \left( \msgf{V}{X} - \msgf{V}{X} \dW_X \msgf{V}{X} \right) \msgf{\xi}{X}
       + \msgf{V}{X} \dW_X \msgb{V}{X} \msgb{\xi}{X}
       \IEEEeqnarraynumspace\label{eqn:ProofTabSingleEdgeGMPm3}\\
 & = & \msgf{m}{X} - \msgf{V}{X} \dW_X \left( \msgf{m}{X} - \msgb{m}{X} \right) \\
 & = & \msgf{m}{X} - \msgf{V}{X} \dxi_X,
\end{IEEEeqnarray}
and (\ref{eqn:TabSingleEdgeGMPdxi}) follows by multiplication with $\msgf{W}{X}$.
\end{proofof}

\begin{proofof}{of (\ref{eqn:TabObsGMPE})}
From (\ref{eqn:TabEquGMPW}) and (\ref{eqn:TabMatMultGMPWb}), we have 
\begin{equation}
\msgf{W}{Z} = \msgf{W}{X} + A^\T \msgb{W}{Y} A,
\end{equation}
from which we obtain
\begin{IEEEeqnarray}{rCl}
\msgf{W}{X} & = & \msgf{W}{Z} - A^\T \msgb{W}{Y} A \\
 & = & \msgf{W}{Z} F.   \label{eqn:proofObsGMPWXWZF}
\IEEEeqnarraynumspace
\end{IEEEeqnarray}
Thus $\msgf{V}{Z}\msgf{W}{X} = F$ and 
\begin{equation}  \label{eqn:msgfVZFmsgfVX}
\msgf{V}{Z} = F \msgf{V}{X}.
\end{equation}
On the other hand, we have 
\begin{equation}
\msgf{V}{Z} = \big( I - \msgf{V}{X}  A^\T G A \big) \msgf{V}{X}
\end{equation}
from (\ref{eqn:TabObsGMPVf}),
and $F = I - \msgf{V}{X}  A^\T G A$ follows.
\end{proofof}

\begin{proofof}{of (\ref{eqn:TabObsGMPdxiZ})}
Using (\ref{eqn:TabEquGMPdxi}), (\ref{eqn:TabMatMultGMPdxi}), 
and (\ref{eqn:TabSingleEdgeGMPdxi2}), we have
\begin{IEEEeqnarray}{rCl}
\dxi_X & = & \dxi_Z + A^\T \dxi_Y \\
  & = & \dxi_Z + A^\T \big( \msgb{W}{Y} m_Y - \msgb{W}{Y} \msgb{m}{Y} \big).
    \IEEEeqnarraynumspace\label{eqn:ProofTabObsGMPdxiZ1rev}
\end{IEEEeqnarray}
Using (\ref{eqn:TabMatMultGMPm}) and (\ref{eqn:TabSingleEdgeGMPm}), 
we further have
\begin{IEEEeqnarray}{rCl}
m_Y & = & A m_Z \\
  & = & A \big( \msgf{m}{Z} - \msgf{V}{Z} \dxi_Z \big),
        \label{eqn:ProofTabObsGMPdxiZ2rev}
\end{IEEEeqnarray}
and inserting (\ref{eqn:ProofTabObsGMPdxiZ2rev}) into (\ref{eqn:ProofTabObsGMPdxiZ1rev}) 
yields (\ref{eqn:TabObsGMPdxiZ}).
\end{proofof}

\begin{proofof}{of (\ref{eqn:TabObsGMPdxi})}
We begin with $m_X=m_Z$. Using (\ref{eqn:TabSingleEdgeGMPm}), we have
\begin{IEEEeqnarray}{rCl}
\msgf{m}{X} - \msgf{V}{X} \dxi_X
 & = &  \msgf{m}{Z} - \msgf{V}{Z} \dxi_Z \\
 \IEEEeqnarraymulticol{3}{l}{
 \quad = \msgf{m}{X} + \msgf{V}{X} A^\T G (\msgb{m}{Y} - A\msgf{m}{X}) -\msgf{V}{X} F^\T \dxi_Z,
  }
  \IEEEeqnarraynumspace
\end{IEEEeqnarray}
where the second step uses (\ref{eqn:TabObsGMPmf}) and 
$\msgf{V}{Z} = F \msgf{V}{X} = (F \msgf{V}{X})^\T$
from (\ref{eqn:msgfVZFmsgfVX}).
Subtracting $\msgf{m}{X}$ and multiplying by $\msgf{V}{X}^{-1}$ yields (\ref{eqn:TabObsGMPdxi}).
\end{proofof}

\begin{proofof}{of (\ref{eqn:TabObsGMPdW})}
We begin with $V_X=V_Z$. Using (\ref{eqn:TabSingleEdgeGMPV2}), we have
\begin{IEEEeqnarray}{rCl}
\msgf{V}{X} - \msgf{V}{X} \dW_X \msgf{V}{X}
 & = & \msgf{V}{Z} - \msgf{V}{Z} \dW_Z \msgf{V}{Z} \\
 \IEEEeqnarraymulticol{3}{l}{
 \quad = \msgf{V}{X} - \msgf{V}{X} A^\T G A \msgf{V}{X} - \msgf{V}{X} F^\T \dW_Z F \msgf{V}{X},
  }
  \IEEEeqnarraynumspace
\end{IEEEeqnarray}
where the second step uses (\ref{eqn:TabObsGMPVZ}) and (\ref{eqn:msgfVZFmsgfVX}).
Subtracting $\msgf{V}{X}$ and multiplying by $\msgf{V}{X}^{-1}$
yields (\ref{eqn:TabObsGMPdW}).
\end{proofof}

\begin{proofof}{of (\ref{eqn:TabObsGMPdWZ})}
As we have already established (\ref{eqn:TabObsGMPdW}), 
we only need to prove 
\begin{equation} \label{eqn:proofAGAeqAWYAF}
A^\T G A = A^\T \msgb{W}{Y} AF.
\end{equation}
Using (\ref{eqn:TabObsGMPE}), we have
\begin{IEEEeqnarray}{rCl}
A^\T G A 
 & = & \msgf{W}{X} (I - F) \\
 & = & \msgf{W}{X} \msgf{V}{Z} A^\T \msgb{W}{Y} A
       \IEEEeqnarraynumspace\\
 & = & A^\T \msgb{W}{Y} A \msgf{V}{Z} \msgf{W}{X},
\end{IEEEeqnarray}
where the last step follows from $A^\T G A = (A^\T G A)^\T$.
Inserting (\ref{eqn:proofObsGMPWXWZF}) then yields (\ref{eqn:proofAGAeqAWYAF}).
\end{proofof}

\begin{proofof}{of (\ref{eqn:TabInpGMPdF2})}
From (\ref{eqn:TabAddGMPVf}) and (\ref{eqn:TabMatMultGMPVf}), we have 
\begin{equation}
\msgf{V}{Z} = \msgf{V}{X} + A \msgf{V}{Y} A^\T,
\end{equation}
from which we obtain
\begin{IEEEeqnarray}{rCl}
\msgf{V}{X} & = & \msgf{V}{Z} - A \msgf{V}{Y} A^\T \\
  & = & \msgf{V}{Z} \dF.
 \IEEEeqnarraynumspace
\end{IEEEeqnarray}
Thus $\msgf{W}{Z} \msgf{V}{X} = \dF$ and
\begin{equation}  \label{eqn:proofInpVXVZdF}
\msgf{W}{Z} = \dF \msgf{W}{X}.
\end{equation}
On the other hand, we have 
\begin{equation}
\msgf{W}{Z} = \big( I - \msgf{W}{X} AHA^\T \big) \msgf{W}{X}
\end{equation}
from (\ref{eqn:InpGMPWZ}),
and $\dF = I - \msgf{W}{X} AHA^\T$ follows.
\end{proofof}

\begin{proofof}{of (\ref{eqn:TabInpGMPmXZ})}
Using (\ref{eqn:TabAddGMPmb}), (\ref{eqn:TabMatMultGMPm}), 
and (\ref{eqn:TabSingleEdgeGMPm}), we have
\begin{IEEEeqnarray}{rCl}
m_X & = & m_Z - A m_Y \\
 & = & m_Z - A \big( \msgf{m}{Y} - \msgf{V}{Y} \dxi_Y \big).
  \label{eqn:proofInpmX1}
 \IEEEeqnarraynumspace
\end{IEEEeqnarray}
Using (\ref{eqn:TabAddGMPdxi}), (\ref{eqn:TabMatMultGMPdxi}),
and (\ref{eqn:TabSingleEdgeGMPdxi}), we further have
\begin{IEEEeqnarray}{rCl}
\dxi_Y & = & A^\T \dxi_Z \\
 & = & A^\T \big( \msgf{\xi}{Z} - \msgf{W}{Z} m_Z \big),
       \label{eqn:proofInpdxiY}
 \IEEEeqnarraynumspace
\end{IEEEeqnarray}
and inserting (\ref{eqn:proofInpdxiY}) into (\ref{eqn:proofInpmX1}) 
yields (\ref{eqn:TabInpGMPmXZ}).
\end{proofof}

\begin{proofof}{of (\ref{eqn:TabInpGMPmX})}
We begin with $\dxi_X = \dxi_Z$ from (\ref{eqn:TabAddGMPdxi}).
Using (\ref{eqn:TabSingleEdgeGMPdxi}), we have 
\begin{IEEEeqnarray}{rCl}
\IEEEeqnarraymulticol{3}{l}{
 \msgf{\xi}{X} - \msgf{W}{X} m_X = \msgf{\xi}{Z} - \msgf{W}{Z} m_Z 
 }\\
 \quad & = & \msgf{\xi}{X} + \msgf{W}{X} AH (\msgf{\xi}{Y} - A^\T \msgf{\xi}{X}) - \msgf{W}{X} \dF^\T m_Z, 
 \IEEEeqnarraynumspace
\end{IEEEeqnarray}
where the second step uses (\ref{eqn:InpGMPxiZ}) and 
$\msgf{W}{Z} = (\dF \msgf{W}{X})^\T$ 
from (\ref{eqn:proofInpVXVZdF}).
Subtracting $\msgf{\xi}{X}$ and multiplying by $\msgf{V}{X}$ yields (\ref{eqn:TabInpGMPmX}).
\end{proofof}

\begin{proofof}{ of (\ref{eqn:TabInpGMPVX})}
We begin with $\dW_X = \dW_Z$ from (\ref{eqn:TabAddGMPdW}). 
Using (\ref{eqn:TabSingleEdgeGMPdW}), we have 
\begin{IEEEeqnarray}{rCl}
\IEEEeqnarraymulticol{3}{l}{
 \msgf{W}{X} - \msgf{W}{X} V_X \msgf{W}{X} = \msgf{W}{Z} - \msgf{W}{Z} V_Z \msgf{W}{Z} 
 }\\
 \quad & = & \msgf{W}{X} - \msgf{W}{X} AHA^\T \msgf{W}{X} - \msgf{W}{X} \dF^\T V_Z \dF \msgf{W}{X},
 \IEEEeqnarraynumspace
\end{IEEEeqnarray}
where the second step uses (\ref{eqn:InpGMPWZ}) and (\ref{eqn:proofInpVXVZdF}).
Subtracting $\msgf{W}{X}$ and multiplying by $\msgf{V}{X}$ yields (\ref{eqn:TabInpGMPVX}).
\end{proofof}

\begin{proofof}{of (\ref{eqn:TabInpGMPVXZ})}
Since we have already established (\ref{eqn:TabInpGMPVX}), 
we only need to prove
\begin{equation} \label{eqn:proofAHATAVYATdF}
A H A^\T = A \msgf{V}{Y} A^\T \dF.
\end{equation}
Using (\ref{eqn:TabInpGMPdF2}), we have 
\begin{IEEEeqnarray}{rCl}
A H A^\T & = & \msgf{V}{X} (I - \dF) \\
 & = & \msgf{V}{X} \msgf{W}{Z} A \msgf{V}{Y} A^\T \\
 & = & A \msgf{V}{Y} A^\T \msgf{W}{Z} \msgf{V}{X},
 \IEEEeqnarraynumspace
\end{IEEEeqnarray}
where the last step follows from $A H A^\T = (A H A^\T)^\T$.
Inserting (\ref{eqn:proofInpVXVZdF}) then yields (\ref{eqn:proofAHATAVYATdF}).
\end{proofof}

\section{Message Passing in \Fig{fig:FGCS} and Proofs}
\label{appsec:LeastSquaresMessagePassing}

In this appendix, we demonstrate how all the quantities 
pertaining to computations mentioned in Sections \ref{sec:LeastSquares} and~\ref{sec:VarEst},
as well as the proof of the theorem in Section~\ref{sec:LeastSquares},
are obtained by symbolic message passing using the tables in Appendix~\ref{appsec:GMPTables}.
The key ideas of this section are from \cite[Section V.C]{LDHKLK:fgsp2007}.

Throughout this section,
$\sigma_1,\ldots,\sigma_K$ are fixed.

\subsection{Key Quantities $\dxi_{X_k}$ and $\dW_{X_k}$}

The pivotal quantities of this section are
the dual mean vector $\dxi_{\tilde U_k}$ and the dual precision matrix $\dW_{\tilde U_k}$.
Concerning the former, we have
\begin{IEEEeqnarray}{rCl}
\dxi_{\tilde U_k} & = & \dxi_{X_k} = \dxi_{X_0} = \dxi_{Y} 
       \IEEEeqnarraynumspace\label{eqn:LeastSquaresMPdxi1}\\
 & = & -\dW_Y y,  \label{eqn:LeastSquaresMPdxi2}
\end{IEEEeqnarray}
for $k=1,\ldots, K,$
where (\ref{eqn:LeastSquaresMPdxi1}) follows from (\ref{eqn:TabAddGMPdxi}),
and (\ref{eqn:LeastSquaresMPdxi2}) follows from 
\begin{IEEEeqnarray}{rCl}
\dxi_{Y} & = & \dW_{Y} (\msgf{m}{Y} - \msgb{m}{Y}) \\
 & = & -\dW_{Y}y
\end{IEEEeqnarray}
since $\msgf{m}{Y}=0$.

Concerning $\dW_{X_k}$, we have
\begin{IEEEeqnarray}{rCl}
\dW_{\tilde U_k} & = & \dW_{X_k} = \dW_{X_0} = \dW_Y 
        \IEEEeqnarraynumspace\label{eqn:LeastSquaresMPdW1}\\
 & = & \dW \text{~as defined in (\ref{eqn:LeastSquarestildeW})}
       \IEEEeqnarraynumspace\label{eqn:LeastSquaresMPdW2}
\end{IEEEeqnarray}
for $k=1,\ldots, K,$
where (\ref{eqn:LeastSquaresMPdW1}) follows from (\ref{eqn:TabAddGMPdW}),
and (\ref{eqn:LeastSquaresMPdW2}) follows from 
\begin{equation}
\dW_Y  = \big( \msgf{V}{Y} + \msgb{V}{Y} \big)^{-1}
\end{equation}
with $\msgb{V}{Y}=0$ and 
\begin{equation} \label{eqn:LeastSqaresMsgfVY}
\msgf{V}{Y} = \sum_{k=1}^K \sigma_k^2 b_k b_k^\T + \sigma^2 I.
\end{equation}

The matrix $\dW$ can be computed without matrix inversion as follows.
First, we note that
\begin{IEEEeqnarray}{rCl}
\dW_{X_0} & = & \big( \msgf{V}{X_0} + \msgb{V}{X_0} \big)^{-1} \\
  & = & \big( 0 + \msgb{V}{X_0} \big)^{-1} \\
  & = & \msgb{W}{X_0}.
\end{IEEEeqnarray}
Second, using (\ref{eqn:InpGMPWZ}), 
the matrix $\msgb{W}{X_0}$ can be computed by the backward recursion
\begin{equation}
\msgb{W}{X_{k-1}} = \msgb{W}{X_{k}} - (\msgb{W}{X_k} b_k)
           (\sigma_k^{-2} + b_k^\T \msgb{W}{X_{k}} b_k)^{-1} (\msgb{W}{X_k} b_k)^\T
\end{equation}
starting from $\msgb{W}{X_K} = \sigma^{-2} I$.
The complexity of this alternative computation of $\dW$ is 
$O(n^2 K)$; by contrast, the direct computation of (\ref{eqn:LeastSquarestildeW})
(using Gauss-Jordan elimination for the matrix inversion) has complexity $O(n^2 K + n^3)$.

\subsection{Posterior Distribution and MAP estimate of $U_k$}

For fixed $\sigma_1,\ldots,\sigma_K,$
the MAP estimate of $U_k$ is the mean $m_{U_k}$ of the (Gaussian) posterior 
of $U_k$. From (\ref{eqn:TabSingleEdgeGMPm}) and (\ref{eqn:TabMatMultGMPdxi}), we have
\begin{IEEEeqnarray}{rCl}
m_{U_k} & = & \msgf{m}{U_k} - \msgf{V}{U_k} \dxi_{U_k} 
          \IEEEeqnarraynumspace\\
 & = & 0 - \sigma_k^2 b_k^\T \dxi_{\tilde U_k},
\end{IEEEeqnarray}
and (\ref{eqn:LeastSquaresMPdxi2}) yields
\begin{equation} \label{eqn:LeastSquaresPosteriorMean}
m_{U_k} = \sigma_k^2 b_k^\T \dW y,
\end{equation}
which proves (\ref{eqn:MAPEstimate}).

For re-estimating the variance $\sigma_k^2$ as in Section~\ref{sec:VarEst},
we also need the variance $\sigma_{U_k}^2$ of the posterior distribution of $U_k$. 
From (\ref{eqn:TabSingleEdgeGMPV2}) and (\ref{eqn:TabMatMultGMPdW}), we have
\begin{IEEEeqnarray}{rCl}
\sigma_{U_k}^2 & = & \msgf{V}{U_k} - \msgf{V}{U_k} \dW_{U_k} \msgf{V}{U_k} 
         \IEEEeqnarraynumspace\\
 & = & \sigma_k^2 - \sigma_k^2 b_k^\T \dW_{\tilde U_k} b_k \sigma_k^2,
         \IEEEeqnarraynumspace
\end{IEEEeqnarray}
and (\ref{eqn:LeastSquaresMPdW2}) yields
\begin{equation} \label{eqn:LeastSquaresPosteriorVar}
\sigma_{U_k}^2 = \sigma_k^2 - \sigma_k^2 b_k^\T \dW b_k \sigma_k^2.
\end{equation}

\subsection{Likelihood Function and Backward Message of $U_k$}

We now consider
the backward message along the edge $U_k$,
which is the likelihood function $p(y \cond u_k, \sigma_1,\ldots,\sigma_K)$,
for fixed $y$ and fixed $\sigma_1,\ldots,\sigma_K$, up to a scale factor. 
For use in Section~\ref{sec:ProofTheorem} below,
we give two different expressions both for the mean $\msgb{m}{U_k}$
and for the variance $\msgb{\sigma}{U_k}^2$ of this message.

As to the latter, we have 
\begin{equation}
\msgb{W}{U_k} = b_k^\T \msgb{W}{\tilde U_k} b_k
\end{equation}
from (\ref{eqn:TabMatMultGMPWb}), and thus
\begin{equation}  \label{eqn:LeastSquaresMsgSigmaUkb}
\msgb{\sigma}{U_k}^2 = \big( b_k^\T \msgb{W}{\tilde U_k} b_k \big)^{-1}.
\end{equation}
We also note (from (\ref{eqn:TabAddGMPVb})) that
\begin{IEEEeqnarray}{rCl}
\msgb{W}{\tilde U_k} & = & \big( \msgf{V}{X_{k-1}} + \msgb{V}{X_k} \big)^{-1}
     \IEEEeqnarraynumspace\\
 & = & W_k \text{~as defined in (\ref{eqn:DefWk}).}
       \IEEEeqnarraynumspace\label{eqn:MsgbWtildeUisWk}
\end{IEEEeqnarray}
Alternatively, we have 
\begin{IEEEeqnarray}{rCl}
\msgb{\sigma}{U_k}^2
  & = & \dW_{U_k}^{-1} - \msgf{V}{U_k} \\
  & = & (b_k^\T \dW_{\tilde U_k} b_k)^{-1} - \sigma_k^2 \IEEEeqnarraynumspace\\
  & = & (b_k^\T \dW b_k)^{-1} - \sigma_k^2, \label{eqn:LeastSquaresMsgbSigmadW}
\end{IEEEeqnarray}
where we used (\ref{eqn:TabSingleEdgeGMPdWDef}), (\ref{eqn:TabMatMultGMPdW}),
and (\ref{eqn:LeastSquaresMPdW2}).

As to the mean $\msgb{m}{U_k}$, we have
\begin{IEEEeqnarray}{rCl}
\msgb{\xi}{U_k} & = & b_k^\T \msgb{\xi}{\tilde U_k} \\
 & = & b_k^\T \msgb{W}{\tilde U_k} \msgb{m}{\tilde U_k} \\
 & = & b_k^\T \msgb{W}{\tilde U_k} ( \msgb{m}{X_k} - \msgf{m}{X_{k-1}} )
       \IEEEeqnarraynumspace\\
 & = & b_k^\T \msgb{W}{\tilde U_k} y
\end{IEEEeqnarray}
from (\ref{eqn:TabMatMultGMPxib}) and (\ref{eqn:TabAddGMPmb}),
and thus
\begin{IEEEeqnarray}{rCl}
\msgb{m}{U_k} & = & \msgb{\sigma}{U_k}^2 \msgb{\xi}{U_k} \\
  & = & \big( b_k^\T \msgb{W}{\tilde U_k} b_k \big)^{-1} b_k^\T \msgb{W}{\tilde U_k} y
  \IEEEeqnarraynumspace\label{eqn:LeastSquaresMsgbmU}
\end{IEEEeqnarray}
from (\ref{eqn:LeastSquaresMsgSigmaUkb}).
Alternatively, we have
\begin{IEEEeqnarray}{rCl}
\msgb{m}{U_k} & = & \msgf{m}{U_k} - \dW_{U_k}^{-1} \dxi_{U_k} \IEEEeqnarraynumspace\\
 & = & 0 - (b_k^\T \dW_{\tilde U_k} b_k)^{-1} b_k^\T \dxi_{\tilde U_k} \IEEEeqnarraynumspace\\
 & = & (b_k^\T \dW b_k)^{-1} b_k^\T \dW y, 
       \IEEEeqnarraynumspace\label{eqn:LeastSquaresMsgbmUdW}
\end{IEEEeqnarray}
where we used (\ref{eqn:TabSingleEdgeGMPdxiDef}), (\ref{eqn:TabMatMultGMPdW}), 
(\ref{eqn:TabMatMultGMPdxi}), (\ref{eqn:LeastSquaresMPdxi2}), and (\ref{eqn:LeastSquaresMPdW2}).

\subsection{Proof of the Theorem in Section~\ref{sec:LeastSquares}}
\label{sec:ProofTheorem}

Let $\sigma_1,\ldots,\sigma_K$ be fixed at a local 
maximum or at a saddle point of the likelihood $p(y \cond \sigma_1,\ldots,\sigma_K)$.
Then 
\begin{equation}
\sigma_k = \argmax_{\sigma_k}
    p(y \cond \sigma_1,\ldots,\sigma_K)
\end{equation}
and
\begin{equation} \label{eqn:LeastSquaresProofSigmak}
\sigma_k^2 = \max \{ 0, \msgb{m}{U_k}^2 - \msgb{\sigma}{U_k}^2 \}
\end{equation}
from (\ref{eqn:BasicScalarSigmaML}). 
From (\ref{eqn:LeastSquaresMsgbmU}) and (\ref{eqn:LeastSquaresMsgSigmaUkb}), we have
\begin{equation}
\msgb{m}{U_k}^2 - \msgb{\sigma}{U_k}^2 
= \frac{\big( b_k^\T \msgb{W}{\tilde U_k} y \big)^2}{\big( b_k^\T \msgb{W}{\tilde U_k} b_k \big)^2} 
   - \frac{1}{b_k^\T \msgb{W}{\tilde U_k} b_k}
\end{equation}
With (\ref{eqn:MsgbWtildeUisWk}), it is obvious that 
$\msgb{m}{U_k}^2 - \msgb{\sigma}{U_k}^2 \leq 0$
if and only if (\ref{eqn:LeastSquaresTheoremMsgbUZero}) holds.

As to (\ref{eqn:LeastSquaresTheoremdWIneq}), 
we have
\begin{equation} \label{eqn:LeastSquaresProofmsgbm2sigma}
\msgb{m}{U_k}^2 - \msgb{\sigma}{U_k}^2 = 
  \frac{\big( b_k^\T \dW y \big)^2}{\big( b_k^\T \dW b_k \big)^2}
- \frac{1}{b_k^\T \dW b_k} + \sigma_k^2
\end{equation}
from (\ref{eqn:LeastSquaresMsgbmUdW}) and (\ref{eqn:LeastSquaresMsgbSigmadW}).
We now distinguish two cases. 
If $\sigma_k^2 > 0$, (\ref{eqn:LeastSquaresProofSigmak}) and (\ref{eqn:LeastSquaresProofmsgbm2sigma}) 
together imply 
\begin{equation}
\frac{\big( b_k^\T \dW y \big)^2}{\big( b_k^\T \dW b_k \big)^2}
- \frac{1}{b_k^\T \dW b_k} = 0.
\end{equation}
On the other hand, if $\sigma_k^2 = 0$, 
(\ref{eqn:LeastSquaresProofSigmak}) and (\ref{eqn:LeastSquaresProofmsgbm2sigma})
imply
\begin{equation}
\frac{\big( b_k^\T \dW y \big)^2}{\big( b_k^\T \dW b_k \big)^2}
- \frac{1}{b_k^\T \dW b_k} \leq 0.
\end{equation}
Combining these two cases yields (\ref{eqn:LeastSquaresTheoremdWIneq}).

\newcommand{\COM}{IEEE Trans.\ Communications}
\newcommand{\COMMag}{IEEE Communications Mag.}
\newcommand{\IT}{IEEE Trans.\ Information Theory}
\newcommand{\JSAC}{IEEE J.\ Select.\ Areas in Communications}
\newcommand{\SP}{IEEE Trans.\ Signal Proc.}
\newcommand{\SPMag}{IEEE Signal Proc.\ Mag.}
\newcommand{\JSP}{IEEE J.\ Select.\ Topics in Signal Proc.}
\newcommand{\ProcIEEE}{Proceedings of the IEEE}


\begin{thebibliography}{99}

\bibitem{RoGha:unifLGM1999}
S.\ Roweis and Z.\ Ghahramani,
``A unifying review of linear Gaussian models,''
\emph{Neural Computation,} vol.~11, pp.~305--345, Feb.\ 1999.

\bibitem{KSH:LinEs2000b}
T.~Kailath, A.~H.\ Sayed, and B.~Hassibi,
\emph{Linear Estimation.}
Prentice Hall, NJ, 2000.

\bibitem{DuKa:tsassm}
J.~Durbin and S.~J.~Koopman,
\emph{Time Series Analysis by State Space Methods.}
Oxford Univ.\ Press, 2012.

\bibitem{Bi:PRML}
C.\ Bishop,
\emph{Pattern Recognition and Machine Learning.}
Springer, 2006.


\bibitem{Lg:ifg2004}
H.-A.\ Loeliger,
``An introduction to factor graphs,''
\emph{\SPMag,} Jan.\ 2004, pp.~28--41.

\bibitem{LDHKLK:fgsp2007}
H.-A.~Loeliger, J.~Dauwels, Junli~Hu, S.~Korl, Li~Ping, and F.~R.~Kschischang,
``The factor graph approach to model-based signal processing,''
\emph{\ProcIEEE,} vol.~95, no.~6, pp.~1295--1322, June 2007.


\bibitem{McKay:Bi1992}
D.~J.~C\ MacKay,
``Bayesian interpolation,''
\emph{Neural Comp.,} vol.~4, n.~3, pp.~415--447, 1992.

\bibitem{Gull:bii1988}
S.\ Gull,
``Bayesian inductive inference and maximum entropy,''
in \emph{Maximum-entropy and Bayesian Methods in Science and Engineering,} 
G.~J.\ Erickson and C.~R.\ Smith, eds.,
Kluwer 1988, pp.~53--74.

\bibitem{Neal:blnn1996b}
R.~M.\ Neal,
\emph{Bayesian Learning for Neural Networks,}
New York: Springer Verlag, 1996.

\bibitem{Tip:RVM2001}
M.~E.\ Tipping,
``Sparse Bayesian learning and the relevance vector machine,''
\emph{J.~Machine Learning Research,}
vol.~1, pp.~211--244, 2001.


\bibitem{Tipp:fmlm2003}
M.~E.\ Tipping and A.~C.~Faul,
``Fast marginal likelihood maximisation for sparse Bayesian models,''
\emph{Proc.\ 9th Int.\ Workshop on Artificial Intelligence and Statistics,}
2003.

\bibitem{WiNag:nvARD2007}
D.~Wipf and S.~Nagarajan,
``A new view of automatic relevance determination,''
\emph{Advances in Neural Information Processing Systems,} 
pp.~1625--1632, 2008.

\bibitem{WiRao:splbs2004}
D.~P.\ Wipf and B.~D.\ Rao,
``Sparse Bayesian learning for basis selection,''
\emph{\SP,} vol.~52, no.~8, Aug.\ 2004, pp.~2153--2164.



\bibitem{DLR:EM1977}
A.~P.\ Dempster, N.~M.\ Laird, and D.~B.\ Rubin,
``Maximum likelihood from incomplete data via the EM algorithm,''
\emph{Journal of the Royal Statistical Society,} 
vol.~39, Series~B, pp.~1--38, 1977.


\bibitem{StSe:cmem2004}
P.~Stoica and Y.~Sel\'en,
``Cyclic minimizers, majorization techniques, 
and the expectation-maximization algorithm: a refresher,''
\emph{\SPMag,} January 2004, pp.~112--114.

\bibitem{GhaHi:pelds1996}
Z.\ Ghahramani and G.~E.\ Hinton,
\emph{Parameter Estimation for Linear Dynamical Systems.}
Techn.\ Report CRG-TR-96-2, Univ.\ of Toronto, 1996.

\bibitem{DEKL:em2013arXiv}
J.~Dauwels, A.~Eckford, S.~Korl, and H.-A.~Loeliger,
``Expectation maximization as message passing---Part~I: principles and Gaussian messages,''
arXiv:0910.2832.



\bibitem{KFL:fg2001}
F.~R.\ Kschischang, B.~J.\ Frey, and H.-A.\ Loeliger,
\emph{``Factor graphs and the sum-product algorithm,''}
\emph{\IT,} vol.~47, pp.~498--519, Feb.\ 2001.


\bibitem{BrLg:sise2014c}
L.\ Bruderer and H.-A.\ Loeliger,
``Estimation of sensor input signals that are neither bandlimited nor sparse,''
\emph{2014 Information Theory \& Applications Workshop (ITA),}
San Diego, CA, Feb.\ 9--14, 2014.

\bibitem{Diss:Bruderer2015}
L.\ Bruderer,
\emph{Input Estimation and Dynamical System Identification: New Algorithms and Results.}
PhD thesis at ETH Zurich No 22575, 2015.

\bibitem{Bi:fmdse}
G.~J.\ Bierman,
\emph{Factorization Methods for Discrete Sequential Estimation.}
New York: Academic Press, 1977.


\bibitem{BrMaLg:ISIT2015c}
L.\ Bruderer, H.\ Malmberg, and H.-A.\ Loeliger,
``Deconvolution of weakly-sparse signals and dynamical-system identification 
by Gaussian message passing,''
\emph{2015 IEEE Int.\ Symp.\ on Information Theory (ISIT),} Hong Kong, June 14--19, 2015.

\bibitem{ZML:spICASSP2016c}
N.~Zalmai, H.~Malmberg, and H.~A. Loeliger,
``Blind deconvolution of sparse but filtered pulses with linear state space models,''
\emph{41th IEEE Int.\ Conf.\ on Acoustics, Speech and Signal Processing (ICASSP),}
Shanghai, China, March 20--25, 2016.


\bibitem{WaLg:OIKSd2016}
F.~Wadehn, L.~Bruderer, V.~Sahdeva, and H.-A.\ Loeliger,
``Outlier-insensitive Kalman smoothing and marginal message passing,''
in preparation.


\bibitem{DMM:mpcs}
D.~L.\ Donoho, A.~Maleki, and A.~Montanari,
``Message-passing algorithms for compressed sensing'',
\emph{Proc.\ National Academy of Sciences,}
vol.~106, no.~45, pp.~18914--18919, 2009.




\end{thebibliography}
\end{document}